\documentclass[12pt,letterpaper,notitlepage]{article}
\usepackage[utf8]{inputenc}
\usepackage[margin=1in]{geometry}
\usepackage{setspace}
\usepackage{enumitem}%
\usepackage{needspace}

\usepackage{amssymb}
\usepackage{amsmath}
\usepackage{amsthm}
\newtheorem{assumption}{Assumption}
\newtheorem{subassumption}{Assumption\normalfont}[assumption]

\newenvironment{assumption*}
 {\ifnum\value{subassumption}=0 \stepcounter{assumption}\fi\subassumption}
 {\endsubassumption}
\newenvironment{assumption+}[1]
 {\renewcommand{\thesubassumption}{#1}\subassumption}
 {\endsubassumption}
 
\newtheorem{claim}{Proposition}
\newtheorem{lemma}{Lemma}
\newtheorem{theorem}{Theorem}

\theoremstyle{definition}

\newtheorem{remark}{Remark}
\usepackage[english]{babel}
\usepackage[authoryear]{natbib}
\usepackage{bbm,graphicx,bm} %
\usepackage[colorlinks = TRUE, allcolors = blue, hypertexnames=false]{hyperref}
\usepackage{float} 
\usepackage{caption}
\usepackage[labelformat=simple,labelfont={}]{subcaption}

\usepackage{booktabs}
\usepackage{graphicx}

\newcommand{\E}{\ensuremath{\mathbb{E}}}

\newcommand{\cv}{\text{cv}}
\newcommand{\Var}{\text{Var}}

\newcommand{\argmin}{\text{argmin}}

\DeclareMathOperator{\maxbias}{\overline{bias}}

\usepackage{mathtools}
\DeclarePairedDelimiter\abs{\lvert}{\rvert}
\DeclarePairedDelimiter\norm{\lVert}{\rVert}

\def\b1{\boldsymbol{1}}

\title{Estimating Treatment Effects Under Bounded Heterogeneity\footnote{We thank Marcella Alsan, Isaiah Andrews, Tim Armstrong, Cl{\'e}ment de Chaisemartin, Ben Deaner, Avi Feller, Phillip Heiler, Peter Hull, Pat Kline, Jing Kong, Matt Masten, Claudia Noack, Jon Roth, Tymon S{\l}oczy\'nski and conference and seminar participants at Emory University, Erasmus
School of
Economics, Econometric Society World Congress, University of Glasgow Adam Smith Business School, UC Berkeley - Statistics, Stanford University, SEA, CEMFI Workshop on Applied Microeconometrics and Panel Data, LMU-NYU Workshop on Advances in Policy Analysis, ASSA, NUS, University of Rochester and University of Bristol 
 for helpful comments and discussion. Liyang Sun gratefully acknowledges support from the Economic and Social Research Council (new investigator grant UKRI607). Yubin Kim, Justin Lee and Tian Xie provided excellent research assistance. An R package
implementing the \texttt{regulaTE} estimator proposed in this paper is available online at \url{https://github.com/lsun20/regulaTEr}.}}
\begin{document}
\author{Soonwoo Kwon\footnote{Department of Economics, Brown University, Email: soonwoo\_kwon@brown.edu} \ and Liyang Sun\footnote{Department of Economics, University College London and CEMFI, Email: liyang.sun@ucl.ac.uk}}

\date{August 2026}

\maketitle
\begin{abstract}

  Specifications that impose constant treatment effects are common and can be
  biased under heterogeneity, whereas fully flexible alternatives can be
  imprecise or infeasible. Under a bound on the target-weighted variance of
  treatment effects, this paper proposes a generalized ridge
  estimator, \texttt{regulaTE}, that yields heterogeneity-aware confidence
  intervals (CIs). The ridge penalty is chosen to minimize the
  heterogeneity-aware CI length in a homoskedastic Gaussian setting; the resulting
  CIs are asymptotically valid even when overlap fails. Varying the bound enables sensitivity
  analysis of departures from constant effects, illustrated in applications to
  unconfoundedness and staggered adoption designs.
\end{abstract}
\smallskip

\noindent\texttt{Key words:} heterogeneous treatment effects, limited overlap, ridge regression

\smallskip

\noindent\texttt{JEL classification codes:} C12, C21, C51.
\newpage

\section{Introduction}
In many empirically relevant settings for causal inference, correctly specifying
a model requires allowing treatment effects to vary flexibly across covariates,
treatment cohorts, or exposure horizons. Estimating the resulting specification
can be imprecise and may be infeasible due to limited overlap when some treatment
effects entering the target average effect are not identified or estimated with low
precision. Consequently, a fully flexible specification may not be desirable
when researchers are primarily interested in a target average, such as the ATE
or ATT, rather than the full distribution of heterogeneous effects. %

For illustration, consider the unconfoundedness design, where treatment is as
good as randomly assigned conditional on the covariates. Many applied studies
estimate a \emph{short regression} that imposes a constant treatment effect:
\begin{equation*}
Y_i = d_i \beta_{\text{short}} +  x_i'\gamma_{\text{short}} + u_i,
\end{equation*}
and interpret the estimated regression coefficient $\hat \beta_{\text{short}}$ as an estimate of the target average effect. 
With heterogeneous effects, this target average interpretation generally requires a more flexible \emph{long regression} specification, which includes treatment-by-covariate
interactions.

The comparison between short and long regressions also arises in other
designs, such as staggered-adoption designs and stratified randomized experiments, although the sources of heterogeneity and limited overlap differ. 
Revisiting \citet{angrist1998},
\citet[Chapter 3.3.1]{angrist2009mostly} argue that if heterogeneity is limited,
the short regression may still be preferable to unbiased but noisier long regression estimators of the average effect. First, when treatment
effects are constant, the short regression is unbiased and most efficient
under homoskedasticity. Second, its bias remains small when treatment effects do
not vary too much. When overlap fails, long regression estimators may be
undefined, whereas the short regression remains feasible by extrapolating
treatment effects to covariate values without overlap. Accordingly, confidence
intervals (CIs) based on the short regression may have coverage close to the
nominal level when heterogeneity is limited.

This paper formalizes that argument for a design-specific target average. Under
a bound $C^2$ on the target-weighted variance of the treatment effects entering
that average (VCATE), we first show how to bias-correct the short regression
CI so that it is heterogeneity-aware and has correct asymptotic coverage. This approach
remains feasible when limited overlap precludes point identification, because
the VCATE bound restricts the unidentified components of the target average.
Varying $C^2$ from zero therefore provides a sensitivity analysis for violations
of constant effects.

Because the short regression estimator does not adjust to the bound, bias-correcting its CI can yield a wide heterogeneity-aware CI when the bound is large. To obtain shorter intervals,
we propose a generalized ridge regression, \texttt{regulaTE}, which coincides
with the short regression estimator when the bound is zero but otherwise can
depart from it. Building on
\citet{armstrong2023biasaware}, we select the ridge penalty to minimize the heterogeneity-aware CI
length under a homoskedastic Gaussian benchmark. In the empirical applications,
the resulting \texttt{regulaTE} CI is shorter than the bias-corrected short regression CI. An
accompanying R package implements the procedure.\footnote{An R package
  implementing the \texttt{regulaTE} estimator and associated sensitivity analysis is available online at
  \url{https://github.com/lsun20/regulaTEr}.}%

We treat $C^{2}$ as a sensitivity parameter rather than a quantity to be
estimated.  Researchers vary $C^{2}$ over a user-specified range. A useful
summary is the ``breakdown value,'' defined as the smallest value at which the
CI includes zero \citep{kline2013sensitivity,masten2020inference}.  Because
$C^2$ bounds VCATE, a familiar measure of dispersion that is also used in
variance decomposition and welfare analyses
\citep{kline2020leave,sanchezbecerra2023robust,dechaisemartin2024estimatingtreatmenteffectheterogeneitysites},
researchers can use substantive knowledge about treatment-effect magnitudes to
specify reference values. In the two empirical applications, subgroup analyses
from the original studies help inform plausible ranges for $C^2$. Data-driven
choice of $C^{2}$ generally leads either to invalid CIs or excessively wide
intervals, as implied by \citet{armstrong_optimal_2018}.

\medskip
\noindent\textbf{Related literature. } First, since \citet{angrist1998}, a
growing literature has characterized short-regression bias under treatment
effect heterogeneity
\citep{gibbons_serrato_urbancic_jem2018,sloczynski_interpreting_2022,goldsmithpinkham2024contamination,double_fe,goodman2021difference}. This
paper instead constructs heterogeneity-aware CIs under an explicit VCATE
bound. Second, several papers estimate or bound average effects under
restrictions on outcomes or treatment effects, including bounded outcomes
\citep{lechner2008note,lee2021bounding}, sparsity
\citep{athey_approximate_2018}, smoothness \citep{armstrong2021finite}, effect
bounds \citep{dechaisemartin2021tradingoff}, and restrictions on outcome changes
in difference-in-differences \citep{manski_how_2018}. The restriction used here
is tied to treatment-effect variance, without imposing a functional form. Third,
inverse-probability-weighted (IPW) estimators remain unbiased under unrestricted
heterogeneity but require strong overlap; with extreme propensity scores, they
can be unstable or asymptotically nonnormal
\citep{rothe2017robust,heiler2021valid}. Trimming improves stability but changes
the target population
\citep{crump_dealing_2009,ma2020robust,sasaki2022estimation}. The proposed CIs
retain the target population and permit inference under limited
overlap. Finally, we adapt the bias-aware inference framework for regularized
linear models developed by \citet{armstrong2023biasaware}, which builds on the
optimal-inference analysis of \citet{armstrong_optimal_2018}, to design-specific
target average effects under a VCATE bound and limited overlap.

Section~\ref{sec:setup} introduces the setup and VCATE restriction. Section~\ref{sec:proposed-method} develops the inference method. Section~\ref{sec:sensitivity} reports sensitivity analyses and applications. Section~\ref{sec:conclusion} concludes.

\section{Setup}\label{sec:setup}

To present the common structure, consider a fixed-design outcome model
\begin{equation}
   \text{long regression: }Y = W\gamma+D\beta+(D \circ \tilde{X}_{-1})\delta +\varepsilon   \label{eq: reg model fs}
\end{equation}
where $W$ contains controls for confounding, $D$ is the treatment vector, and
$\beta$ is the target average effect. The matrix $X$ contains the variables that
index treatment-effect heterogeneity. Depending on the design, $X$ may coincide with $W$.
The matrix $\tilde{X}_{-1}$ is centered with respect to the target weights and
omits the intercept, so $D \circ \tilde{X}_{-1}$ contains treatment interactions
whose coefficients $\delta$ are deviations from the target average effect.

Under the identifying assumptions of each application, \eqref{eq: reg model fs}
is correctly specified after mapping $W$ and $X$ to the design. Because the
design matrices $W$, $D$, and $X$ are treated as fixed, correct specification
means $\mathbb{E}[\varepsilon]= \mathbf{0}$. We call \eqref{eq: reg model fs}
the \emph{long regression}. In practice, researchers often estimate the
\emph{short regression} that restricts $\delta=0$, because the long regression
can be imprecise or fail to be point-identified.

The model also applies to stratified randomized experiments. In that setting,
$W$ may contain stratum fixed effects, and $X$ may contain target-centered
stratum indicators. Random assignment within strata identifies stratum-specific
average effects in the saturated long regression, so no linearity assumption on
potential outcomes is required. The VCATE bound restricts the target-weighted
variance of these effects. The same construction applies more generally to
multi-site randomized experiments with few units per site
\citep{dechaisemartin2024estimatingtreatmenteffectheterogeneitysites} and to
nonseparable panel data models approximated by linear specifications with
individual specific coefficients \citep{chernozhukov2025fisher}.
The following subsections map the common model to staggered adoption and
unconfoundedness designs.

\subsection{Staggered adoption design}\label{sec:staggered-design}

Consider a balanced panel of $n$ units observed in periods
$\mathcal T:=\{0,\ldots,T\}$, with outcome and binary treatment status
\[
\left\{ \{Y_{it}, d_{it}\}_{t=0}^{T}\right\}_{i=1}^n .
\]
For treated units, let
$e_i := \min\{ t : d_{it} = 1 \}$ be the first treatment period, and set
$e_i=\infty$ for never-treated units. Let $\mathcal{E}$ denote the set of finite
first-treatment periods. Cohort $e$ consists of units with $e_i=e$.

Under standard assumptions of parallel trends and no anticipation, the observed
outcome satisfies
\begin{equation}
  Y_{it} = \alpha_{e_i} + \lambda_t + \sum_{\ell \geq 0, \, e \in \mathcal{E}} \tau_{e,\ell} \cdot d_{it} \cdot \mathbf{1}\{ e_i = e ,  t - e_i = \ell \} + \varepsilon_{it},
\label{eq:saturated TWFE}
\end{equation} where $\alpha_{e_i}$ and $\lambda_t$ are cohort and time fixed
effects that account for confounding. Specifically,
$\tau_{e,\ell} = \E[ Y_{i,e+\ell}(1) - Y_{i,e+\ell}(0) \mid e_i = e ]$
is the CATT for cohort $e$ at relative time $\ell$ and may vary freely across
cohorts and relative times.

The estimand considered in this setting is the average effect over treated
cohorts (ATT), defined as
\[
\text{ATT} = \frac{
 \sum_{t \in \mathcal{T}} \sum_{e \in \mathcal{E},e\leq t} \mathbb{P}_n\{d_{it}=1\}\cdot\mathbb{P}_n\{e_i = e \mid d_{it} = 1 \} \cdot \tau_{e,t-e}
}
{ \sum_{t \in \mathcal{T}}  \mathbb{P}_n\{d_{it}=1\},
}
\]
where $\mathbb{P}_n$ denotes empirical frequencies. Thus, the ATT averages
$\tau_{e,t-e}$ using the empirical distribution of treated observations across
cohort-time cells. Following \citet{wooldridge2025twfe},
\eqref{eq:saturated TWFE} can be reparameterized as the ``long'' regression:
\[
Y_{it} = \alpha_{e_i} + \lambda_t +  d_{it} \beta^{\text{ATT}}_{\text{long}}  +   \left( d_{it} \cdot \tilde{x}_{it} \right)'\delta  + \varepsilon_{it}
\]
where $\tilde{x}_{it}$ collects all but one of the centered cohort and relative time indicators for $\ell \geq 0$:%
\[
\tilde{x}_{it} =\begin{pmatrix}
\vdots \\[4pt]
\mathbf{1}\{ e_i = e,\, t - e_i = \ell \}
 - 
\dfrac{
 \sum_{t \in \mathcal{T}} \mathbb{P}_n\{d_{it}=1\}
 \mathbb{P}_n\{e_i = e \mid d_{it} = 1\}
 \mathbf{1}\{t-e=\ell\}
}{
 \sum_{t \in \mathcal{T}} \mathbb{P}_n\{d_{it}=1\}
} \\[6pt]
\vdots
\end{pmatrix}.
\]
Here $W$ contains the cohort and time fixed effects, $D$ collects $d_{it}$, and
$\tilde{X}_{-1}$ collects the row vectors $\tilde{x}_{it}'$.
Heterogeneity-robust difference-in-differences (DiD) estimators of the ATT
\citep{callaway_santanna_ssrn2018,sun_estimating_2020} aggregate cell-specific
estimators $\hat{\tau}_{e,t-e}$ and therefore require an untreated or
not-yet-treated comparison group for each relevant cell. Without further
assumptions, they cannot recover the full ATT if all units are eventually treated,
because the required counterfactual outcomes in later periods are not identified
due to a lack of untreated or not-yet-treated comparisons.
This is the staggered-adoption analogue of limited overlap: later treated cells
lack the untreated or not-yet-treated observations needed to identify the
corresponding counterfactual trends.

Researchers often replace the long regression with the simpler ``static''
two-way fixed effects (TWFE) regression
\begin{equation}
   Y_{it} = \alpha_{e_i} + \lambda_t + \beta_{\text{short}} \cdot d_{it} + \varepsilon_{it}
\label{eq:static TWFE} 
\end{equation}
which omits interactions between $d_{it}$ and cohort and relative-time
indicators. In \eqref{eq: reg model fs}, this restricts $\delta = 0$.
The static specification therefore implicitly assumes constant effects across
cohorts and relative times.
\citet{double_fe} and \citet{goodman2021difference} show that, when effects
vary, this short regression can be biased for the ATT because it may place
negative weights on some cohort-by-event-time effects. Thus, staggered adoption
has the omitted-heterogeneity problem represented by \eqref{eq: reg model fs}.
The static regression remains estimable in such cases because it omits the
unidentified interactions, but its ATT interpretation then relies on restrictions
on heterogeneity.

\subsection{Unconfoundedness design}
 
Consider the unconfoundedness design with $n$ units. For each unit, the
observed data consist of a binary treatment indicator $d_i\in\{0,1\}$, a vector
of confounders $x_i\in\mathbb R^{k+1}$ that includes a constant, and an outcome
$Y_i$. Applied studies frequently combine the
unconfoundedness assumption with linear regression adjustment and estimate the
short specification
\begin{equation}
Y_i = d_i \beta_{\text{short}} +  x_i'\gamma_{\text{short}} + u_i. \label{eq:short reg}
\end{equation}
Examples include
\citet{martinez2014role,favara2015credit,michalopoulos2016long,bobonis2016monitoring,xu2018costs}.

Following \citet{imbens2009recent}, we maintain the linear outcome regressions
implicit in this adjustment strategy while relaxing constant effects. The
observed outcome satisfies
\begin{equation}
Y_i = d_i \tau(x_i) + x_i'\gamma + \varepsilon_i, \label{eq:dgp_selection}
\end{equation}
where $\{\varepsilon_i\}_{i=1}^n$ are mutually independent and mean zero. Under
this inherited linearity, the conditional average treatment effect is
$\tau(x_i) = \beta + \tilde{x}_{i,-1}'\delta$, with
$\beta=\beta^{\text{ATE}}=\E_n[\tau(x_i)]=n^{-1}\sum_{i=1}^n \tau(x_i)$.
Here $\E_n$ denotes the empirical mean, and
$\tilde x_{i,-1}=x_{i,-1} - \E_n[{x}_{i,-1}]$ denotes the demeaned
covariates, excluding the constant. Substitution yields \eqref{eq: reg model fs}
with both the control and effect-modifier matrices equal to $X$:
\begin{equation}
Y_i =  d_i \beta^{\text{ATE}}_{\text{long}} + (d_i \tilde x_{i,-1}  )'\delta  + x_i'\gamma + \varepsilon_i,\ \text{where }\tilde x_{i,-1}=x_{i,-1} - \E_n[{x}_{i,-1}]. \label{eq:long reg}
\end{equation}
The ATT and average treatment effect on the untreated (ATU) are handled
analogously by changing the demeaning weights; the
analysis therefore focuses on the ATE.

Limited overlap can arise with mixed covariates, as is common in applied work.
For example, the covariates may include county indicators together with
continuous variables such as income. If some counties contain only treated or
only untreated observations, county-specific treatment interactions cannot be
estimated separately, leaving the long regression not point-identified. The
short regression remains estimable because it imposes
\(\delta=0\), but need not target the ATE. Thus, the unconfoundedness design
also has the structure in \eqref{eq: reg model fs}.

\subsection{Bounding heterogeneity with VCATE}\label{sec:vcate_bound}

The analysis restricts treatment effect heterogeneity through
$\delta'V_x\delta\leq C^2$, where $V_x$ is the target-weighted
second-moment matrix of the centered treatment-effect modifiers. Across designs,
we call $\delta'V_x\delta$ VCATE; $V_x$ depends on the target parameter.

Under staggered adoption, define \[ V_x := \frac{ \sum_t \mathbb P_n\{d_{it}=1\} \E_n[ \widetilde x_{it}\widetilde x_{it}' \mid d_{it}=1 ] }{ \sum_t\mathbb P_n\{d_{it}=1\} }. \] It follows that \begin{align*} \delta'V_x\delta &= \frac{ \sum_t \mathbb P_n\{d_{it}=1\} \E_n[ (\widetilde x_{it}'\delta)^2 \mid d_{it}=1 ] }{ \sum_t\mathbb P_n\{d_{it}=1\} } \notag\\ &= \frac{ \sum_t\sum_{e\in\mathcal E, e\leq t} \mathbb P_n\{d_{it}=1\} \mathbb P_n\{e_i=e\mid d_{it}=1\} \left( \tau_{e,t-e}-\beta^{\mathrm{ATT}} \right)^2 }{ \sum_t\mathbb P_n\{d_{it}=1\} }. \label{eq:vcatt quadratic restriction} \end{align*} Thus, under staggered adoption, $\delta'V_x\delta$ is the ATT-weighted variance of the cohort-by-event-time treatment effects.

Under the unconfoundedness design, define
\(
V_x := \E_n[\tilde{x}_{i,-1}\tilde{x}_{i,-1}'].
\)
It follows that
\begin{equation}
\delta'V_x\delta = \delta'\,\E_n[\tilde{x}_{i,-1}\tilde{x}_{i,-1}']\,\delta  = \E_n\!\left[(\tilde{x}_{i,-1}'\delta)^2\right]
 = \E_n\!\left[(\tau(x_i)-\E_n[\tau(x_i)])^2\right]
 \label{eq:wtd quadratic restriction}
\end{equation}
is the sample variance of the CATE.

Variance is a familiar measure of dispersion and has a direct interpretation in
sensitivity analysis. VCATE also appears in variance decompositions and
distributional analyses, which can provide empirical benchmarks for
calibration.\footnote{See, for example,
\citet{levy2021fundamental}, \citet{sanchezbecerra2023robust},
\citet{kline2020leave}, and
\citet{dechaisemartin2024estimatingtreatmenteffectheterogeneitysites}.}
In addition, Popoviciu's inequality provides a conservative reference point:
\[ \operatorname{Var}_{\omega}(\tau) \leq \frac{1}{4} \left( \max \tau-\min \tau \right)^2, \]
where the extrema are taken over the treatment effects entering the target
average and $\omega$ denotes the associated weighting measure. This inequality relates the VCATE restriction
to the range of treatment effects.

Assume that $V_x$ is invertible. Throughout, results are reported on the standard-deviation scale $C$
rather than the variance scale $C^2$, because $C$ shares the units of the
outcome. 

\section{Heterogeneity-aware confidence intervals}
\label{sec:proposed-method}

This section develops inference methods that explicitly account for
treatment effect heterogeneity.  
The formal statements retain the distinction in Section~\ref{sec:setup} between
the controls $W$ and the treatment-effect modifiers $X$. This notation covers
the staggered adoption and unconfoundedness mappings, as well as the experimental
settings described above.

\subsection{\texttt{regulaTE} procedure}\label{sec:bias-variance}
For the generic analysis in this section, $n$ denotes the number of observations
after stacking the data; in the staggered-adoption design, these are unit-period
observations.
In a fixed-design setting, estimating the target average effect under a VCATE
bound is equivalent to inference on $\beta$ in~\eqref{eq: reg model fs} subject to
$\delta'V_x\delta \le C^2$. Equivalently, the parameter space is
$(\beta,\gamma',\delta')' \in \Theta_C$, where $\gamma\in\mathbb R^p$,
$\delta\in\mathbb R^k$, and
\[
\Theta_C := \{(\beta,\gamma',\delta')' \in \mathbb{R}^{1+p+k}:\ \delta'V_x\delta \le C^2\}.
\]

Let $Y=(Y_1,\ldots,Y_n)'$, let $D=(d_1,\ldots,d_n)'$ be the realized treatment
vector, and let $W=(w_1,\ldots,w_n)'\in\mathbb R^{n\times p}$ and
$X=(x_1,\ldots,x_n)'\in\mathbb R^{n\times(k+1)}$ be the realized control and
effect-modifier matrices. Assume $W'W$ is invertible and
$D'H_WD>0$, where $H_W=I-W(W'W)^{-1}W'$. The analysis focuses on linear
estimators $\hat\beta_a = a'Y$, where $a\in\mathbb{R}^n$ is nonrandom. Since
$(D,W,X)$ are treated as fixed, the weights $a$ may depend on them. This class
includes standard estimators such as the difference in means, the short
regression, and the long regression when it is defined.

From~\eqref{eq: reg model fs}, the worst-case
bias of $\hat\beta_a$ over $\Theta_C$ is
\begin{equation}\label{eq:worst-case bias}
\sup_{\Theta_C}|\E[\hat\beta_a]-\beta| :=\maxbias_{a,C}
=
\sup_{(\beta,\gamma',\delta')\in\Theta_C}
\left\{a'(D\beta + W\gamma + (D\circ\tilde X_{-1})\delta) - \beta\right\},
\end{equation}
where $D\circ\tilde X_{-1}$ is the matrix with $i$th row equal to
$d_i\tilde x_{i,-1}'$. Since $\Theta_C$ leaves
$\beta$ and $\gamma$ unrestricted, finite worst-case bias requires $a'D=1$ and
$a'W=0$. The remaining bias is linear in $\delta$, and the symmetry of
$\Theta_C$ makes the one-sided supremum in~\eqref{eq:worst-case bias} equal to
$\sup_{\Theta_C}|\E[\hat\beta_a]-\beta|$. Hence, given the weights $a$, the worst-case bias of the
corresponding estimator can be calculated explicitly. For fixed weights $a$, define the
heterogeneity-aware fixed-length confidence interval (FLCI) as
$\hat\beta_a\pm\chi_{a,C}$, where
$\chi_{a,C}:=\hat V_a^{1/2}\cv_\alpha\!\left(\maxbias_{a,C}/\hat V_a^{1/2}\right)$. Here $\cv_\alpha(B)$ denotes the $(1-\alpha)$ quantile of the folded normal
distribution $|N(B,1)|$ and $\hat V_a$ is a consistent estimator for $\Var(\hat{\beta}_{a})$. Lemma~\ref{thm:FLCI} states its high-level validity.
The folded-normal critical value incorporates the estimator's potential
worst-case bias. Validity does not require point identification of $\beta$:
under the VCATE bound, the interval has uniform asymptotic coverage over
$\Theta_C$, including designs with complete lack of overlap.

\begin{lemma}[High-level validity of a heterogeneity-aware CI]\label{thm:FLCI}

Let $a=a_n$ be nonrandom conditional on the fixed design and satisfy
$a'D=1$ and $a'W=0$. 
Assume $\varepsilon_i$ are independent random variables with zero mean. Suppose that there exist constants $K$ and $\eta>0$ such that for all $i$,
$1/K\leq \E_Q[\varepsilon_i^2]\leq K$ and  
$\E_Q[|\varepsilon_i|^{2+\eta}]\leq K$ 
uniformly over $i$ and $Q\in\mathcal Q_n$, and that 
\[
\operatorname{Lind}(a)
:=
\frac{\max_{1\leq i\leq n}a_i^2}
{\sum_{j=1}^n a_j^2}
\longrightarrow 0.
\]
Suppose further that $\widehat V_a$ is a uniformly consistent variance estimator
such that $\widehat V_a/\Var(\hat{\beta}_{a})\rightarrow_p 1$. Then
\[
\liminf_{n\to\infty} \inf_{(\beta, \gamma', \delta') \in \Theta_C} \inf_{Q\in\mathcal Q_n} \mathbb{P}_{Q} 
\left\{
\beta\in
\left[
\widehat\beta_a
\pm
\widehat V_a^{1/2}
\operatorname{cv}_\alpha
\left(
\frac{\maxbias_{a,C}}
{\widehat V_a^{1/2}}
\right)
\right]
\right\}
\geq 1-\alpha,
\]
 where $\cv_\alpha(B)$ denotes the $(1-\alpha)$ quantile of the
folded normal distribution $|N(B,1)|$.
\end{lemma}
\begin{proof}
Write $\hat{\beta}_a-\beta = \E[a'Y]-\beta + a'\varepsilon$ and note that
$\Var(\hat{\beta}_{a})=\Var(a'\varepsilon)=\sum_{i=1}^n a_i^2 \E_Q[\varepsilon_i^2]$ in the fixed-design model.
The uniform moment conditions and
$\operatorname{Lind}(a)\to0$ imply, by the
Lindeberg--Feller central limit theorem generalized by Lemma F.1 of \cite{armstrong_optimal_2018} to heteroskedastic data,
$
\frac{a'\varepsilon}{\sqrt{\Var(\hat{\beta}_{a})}}
\xrightarrow{d}N(0,1)
$
uniformly over $Q\in\mathcal Q_n$. By the assumed consistency of the
variance estimator, the feasible studentized statistic has the same asymptotic
distribution as a standard normal variable shifted by
\[
\frac{\E[a'Y]-\beta}{\sqrt{\Var(\hat{\beta}_{a})}},
\]
whose absolute value is bounded by
$\maxbias_{a,C}/\sqrt{\Var(\hat{\beta}_{a})}$.
The folded-normal critical value is weakly increasing in the magnitude
of this shift. At the maximal standardized bias, the folded-normal critical
value yields limiting coverage of at least $1-\alpha$.
\end{proof}

As shown in Lemma~\ref{claim:short bias} in the Appendix, the worst-case bias
of the short regression estimator is
\[
\maxbias_{a_s,C}
=
C\sqrt{
a_s'(D\circ\tilde X_{-1})V_x^{-1}(D\circ\tilde X_{-1})'a_s},
\qquad
a_s=(D'H_WD)^{-1}H_WD.
\]
Because the short-regression weights $a_s$ do not adjust as $C$ increases,
bias-corrected short-regression CIs can become excessively long. Since the
length of an FLCI $\hat\beta_a\pm\chi_{a,C}$ equals $2\chi_{a,C}$ and increases
in both $\maxbias_{a,C}$ and $\Var(\hat\beta_a)$, improving performance for
$C>0$ requires a better bias-variance trade-off.

Define $\hat\beta_\lambda$ by the generalized ridge least-squares problem
\begin{equation}
  \label{eq:gen_ridge}
  \min_{\beta, \gamma, \delta} \frac{1}{n} \norm{Y - D\beta - W\gamma - (D \circ
    \tilde{X}_{-1})\delta}^{2} + \lambda\, \delta'V_{x}\delta
\end{equation}
where $\lambda>0$ is chosen below. Since $\delta'V_x\delta$ equals VCATE, the
penalty shrinks $\delta$ toward the constant-effects model, with the degree of
shrinkage governed by $\lambda$. This connection to inference is not automatic:
penalized outcome regression generally targets prediction rather than the
bias-variance trade-off for $\beta$.\footnote{Under an $\ell_1$ bound on $\delta$, for instance, the
  LASSO-penalized outcome regression does not produce optimal inference for
  $\beta$. The equivalence between the ridge penalty on the outcome regression
  and the optimal inference procedure for $\beta$ is specific to the quadratic
  VCATE constraint, as noted by \citet{li1982minimaxity} and \citet{armstrong2023biasaware}.} As shown in Section~\ref{sec:lambda choice}, under i.i.d. homoskedastic Gaussian errors, the quadratic form of the VCATE restriction
implies that $\hat{\beta}_\lambda$ lies on the bias-variance frontier for every
$\lambda>0$: no linear estimator achieves both smaller variance and smaller
worst-case bias. All frontier estimators can be written in this form.

Lemma~\ref{lem:outcome ridge} shows that the generalized ridge estimator
$\hat{\beta}_{\lambda}$ can be written as
      \begin{equation}\label{eq:opt_sol}
   \hat{\beta}_{\lambda} = a_{\lambda}'Y =  \frac{\tilde{D}_{\lambda}'
  Y}{\tilde{D}_{\lambda}'D},
  \end{equation}
  where the weights
    $a_{\lambda} := \frac{\tilde{D}_{\lambda}}{\tilde{D}_{\lambda}'D}$ and the residuals
    $\tilde{D}_{\lambda} := D - W\pi_{1,\lambda} - (D \circ
    \tilde{X}_{-1})\pi_{2,\lambda}$ are from a penalized treatment
    regression:
    \begin{equation}
      \label{eq:prop_lag}
      \min_{\pi_{1}, \pi_{2}} \frac{1}{n} \norm{D - W\pi_{1} - (D \circ
    \tilde{X}_{-1})\pi_{2}}^{2} + \lambda \pi_{2}'V_{x}\pi_{2}.
    \end{equation}
The parameter $\lambda$ controls the extent to which the treatment-effect
interactions are partialled out. As $\lambda\to0$,
$\hat\beta_\lambda$ coincides with the long-regression coefficient when the long
regression is well defined. As $\lambda\to\infty$, $\hat\beta_\lambda$ converges to the
short-regression estimator. Thus, $\lambda$ interpolates between fully
partialling out the interactions and imposing the constant-effects restriction.
This representation makes clear that $\lambda$ changes the estimating weights,
rather than only the critical value used for inference.

\textbf{Example (Unconfoundedness design with discrete covariates).} This
interpolation is explicit when covariates are discrete. Fix
$\lambda>0$ and suppose $W=X$ is the saturated indicator matrix for a discrete
covariate. Online Appendix A shows that the estimand of
$\hat{\beta}_{\lambda}$ is a weighted average of CATEs:
    \begin{equation}
        \beta_{\lambda}= \E[\hat{\beta}_{\lambda}] = \E_n\left[\frac{(p(x_i)(1-p(x_i)))(1-\widetilde{x}_{i,-1}'\pi_{2, \lambda})}{\E_n\left[(p(x_i)(1-p(x_i)))(1-\widetilde{x}_{i,-1}'\pi_{2, \lambda})\right]}\tau(x_{i})\right],\label{eq:ridge estimand}
    \end{equation}
which further simplifies to
\begin{equation}\label{eq:explicit weights}
\beta_{\lambda} = \E_n\left[\frac{p(x_i)(1-p(x_i))/(p(x_i)(1-p(x_i))+\lambda)}
{\E_n\left[p(x_i)(1-p(x_i))/(p(x_i)(1-p(x_i))+\lambda)\right]}
\tau(x_{i})\right].
\end{equation}
The weight on each unit's CATE is proportional to 
$p(x_i)(1-p(x_i))/(p(x_i)(1-p(x_i))+\lambda) \in [0,1]$, which equals
$1/(1+\lambda/[p(x_i)(1-p(x_i))])$. This factor is close to one when
$p(x_i)(1-p(x_i))$ is large relative to $\lambda$ and close to zero when it is
small relative to $\lambda$. As $\lambda\to 0$, the shrinkage factor approaches
one and the weights converge to cell shares, recovering the ATE when the
long regression is well defined; without overlap, the limit is the long
regression on the trimmed overlap subsample. As $\lambda\to\infty$,
the weights converge to the short regression weights from \citet{angrist1998},
proportional to $p(x)(1-p(x))$. For intermediate $\lambda$, the ridge estimand
overweights well-overlapped cells relative to the ATE, but less so than the
short regression: cells with $p(x)(1-p(x))\gg\lambda$ receive weights close to
their cell shares, whereas cells with $p(x)(1-p(x))\ll\lambda$ are heavily
downweighted. The weights are nonnegative and sum to one.

\subsubsection{Choosing the penalty parameter}\label{sec:lambda choice}
The penalty parameter $\lambda$ is selected to minimize FLCI half-length in the
homoskedastic Gaussian setting.\footnote{An alternative is to minimize least
squares subject to $\delta'V_x\delta \leq C^2$. Its Lagrangian is a generalized
ridge regression, with the multiplier determining $\lambda$. That multiplier is
chosen to optimize fit subject to the heterogeneity constraint, whereas the
objective here is inference on $\beta$. We therefore choose $\lambda$ to
minimize heterogeneity-aware CI length.} By the
same argument as Lemma~\ref{claim:short bias} in the Appendix, the worst-case
bias of $\hat{\beta}_\lambda$ takes the form
\[
\maxbias_{a_{\lambda},C}=C\cdot\sqrt{a_{\lambda}'(D\circ\tilde{X}_{-1})V_{x}^{-1}(D\circ\tilde{X}_{-1})'a_{\lambda}}.
\] Let
$\sigma^2:=\Var(\varepsilon_i)$, so $\Var(\hat\beta_\lambda)=\sigma^2\norm{a_{\lambda}}^2$.
Let $\lambda^{\ast}_{C}$ denote the penalty parameter value that minimizes the half-length of the
corresponding fixed-length CI:
\begin{equation}\label{eq:opt_lam}
 \lambda^{\ast}_{C} := \argmin_{\lambda} \,\, \sigma\norm{a_{\lambda}} \cdot
  \cv_{\alpha}\left(\maxbias_{a_{\lambda},C}/(\sigma \norm{a_{\lambda}})\right),
\end{equation}
and construct the corresponding CI $\hat{\beta}_{\lambda^*_{C}} \pm
\chi_{a_{\lambda^{\ast}_{C}},C}$.  This inference procedure is denoted \texttt{regulaTE}.

The \texttt{regulaTE} CI is the shortest fixed-length CI based on any linear
estimator in the homoskedastic Gaussian setting:
\begin{theorem}\label{thm:ridge frontier}
  The generalized ridge regression estimator $\hat{\beta}_{\lambda}$ solves the
  bias-variance trade-off problem:
  \begin{equation}
  \label{eq:bias-var}
  \min_{a \in \mathbb{R}^{n}} a'a \quad \text{s.t.} \quad \sup_{(\beta,
    \gamma', \delta') \in \Theta_{C}} a'(D\beta + W\gamma + (D \circ
    \tilde{X}_{-1})\delta) - \beta \leq B
  \end{equation}
  with $B = \maxbias_{a_{\lambda}, C}$. Consequently, the
  \textnormal{\texttt{regulaTE}} CI is the shortest fixed-length CI based on
  linear estimators.
\end{theorem}

The first part of the theorem states that $\hat{\beta}_\lambda$ lies on the
bias-variance frontier for every $\lambda>0$: at bias level
$B=\maxbias_{a_\lambda,C}$, no linear estimator has smaller variance. This
follows from the modulus of continuity characterization of optimal inference
\citep{donoho1994statistical,armstrong_optimal_2018}. The ridge family spans the
entire frontier. Since FLCI half-length depends on $a$ only through variance and
worst-case bias, minimizing CI length over all linear estimators therefore
reduces to minimizing over the ridge family in~\eqref{eq:opt_lam}.

We use this result to select $\lambda$. Outside the i.i.d. homoskedastic
Gaussian model, exact optimality need not hold. The benchmark nevertheless
provides a transparent criterion for selecting the weights, and the
\texttt{regulaTE} CI remains shorter than the heterogeneity-aware
short-regression CI in the applications. The procedure follows the standard
practice of selecting weights under homoskedasticity and reporting robust
standard errors, as in OLS.

\subsubsection{Feasible implementation}\label{sec:implementation}

The feasible procedure begins with an initial estimator
$\hat{\theta}^{\mathrm{init}} = (\hat{\beta}^{\mathrm{init}},
\hat{\gamma}^{\mathrm{init}}{}', \hat{\delta}^{\mathrm{init}}{}')'$ for
$\theta = (\beta, \gamma', \delta')'$, obtained either from the long regression
(when defined) or from a cross-validated generalized ridge regression that
penalizes $\delta'V_x\delta$. Define
$\hat{\varepsilon}_{\mathrm{init},i} = Y_{i} -d_i\hat{\beta}^{\mathrm{init}} -
w_{i}'\hat{\gamma}^{\mathrm{init}}-
d_{i}\tilde{x}_{i,-1}'\hat{\delta}^{\mathrm{init}}$, and the corresponding
initial variance estimator
$\hat{\sigma}^2 = \frac{1}{n} \sum_{i=1}^n \hat{\varepsilon}_{\mathrm{init},i}^2.$

Obtain the feasible $\lambda^*_{C}$ by solving the sample analogue of
\eqref{eq:opt_lam} after substituting $\hat{\sigma}$ for
$\sigma$.\footnote{The fixed-weight validity result applies
directly when this pilot variance is nonrandom conditional on the design; the
applications use the corresponding full-sample plug-in implementation.} Using
this tuning parameter, compute the generalized ridge estimator
$\hat{\beta}_{\lambda^*_{C}}$ and form the robust variance estimator
    \[
    \hat{V}_{\lambda^*_{C},\text{rob}} = \sum_{i=1}^n
  a_{\lambda^*_{C},i}^2 \hat{\varepsilon}_{\mathrm{init},i}^2, \quad
   \text{where } a_{\lambda^*_{C}} = \frac{\tilde{D}_{\lambda^*_{C}}}{\tilde{D}_{\lambda^*_{C}}'D}.
\]
The feasible \texttt{regulaTE} CI is $\hat{\beta}_{\lambda^*_{C}} \pm
\cv_\alpha \left(
  \maxbias_{a_{\lambda^*_{C}},C}/\hat{V}_{\lambda^*_{C},\text{rob}}^{1/2} \right)
\cdot \hat{V}_{\lambda^*_{C},\text{rob}}^{1/2}$. Cluster-robust versions follow analogously.

Lemma~\ref{thm:FLCI} gives high-level validity conditions for fixed weights. To
apply it to $\hat{\beta}_{\lambda^*_{C}}$,
Appendix~\ref{sec:feasible CI primitive condition} verifies consistency of the
robust variance estimator if the maximal Lindeberg weight
  shrinks sufficiently quickly relative to the initial-estimator error. The
  Lindeberg weights can be computed in any application and provide diagnostics
  for the reliability of the normal approximation; see \citet{noack2024bias}
  for an analogous discussion in fuzzy regression discontinuity designs. The
  companion R package reports the maximal Lindeberg weight and warns when it is
  large. Limited overlap may slow convergence below the parametric rate of
  $n^{-1/2}$.

\subsection{Necessity of a VCATE bound}
\label{sec:nec_bd_vcate}
Given a VCATE bound, the previous sections construct \texttt{regulaTE} CIs.
Two alternatives are a CI valid under unrestricted heterogeneity and an
adaptive CI that shrinks with true VCATE while covering a broader class. Both
are infeasible, so an \textit{a priori} bound is necessary.

The no-overlap illustration in this subsection specializes to the
unconfoundedness design with $W=X$ and saturated binary covariates.

Without a parameter-space restriction, the worst-case bias of any linear
estimator is unbounded when overlap fails: the data contain no information about
treatment effects for strata with only treated or untreated units. Formally,
recall from~\eqref{eq:worst-case bias} that the bias is linear in the
parameters. Hence, only unbiased estimators have bounded bias, equal to zero,
when the parameter space is unrestricted. Equation~\eqref{eq:worst-case bias}
implies that
unbiasedness requires $a'D =1$, $a'X = 0$ and $a'(D \circ \tilde{X}_{-1}) =
0$. Suppose overlap fails for the binary $j$th covariate so that $X_{j+1} \leq D$,
where $\leq$ is interpreted elementwise. Writing
$\bar{x}_{j+1} := \E_{n}[x_{i, j+1}]\neq 0$, the conditions for unbiasedness
imply $a'D = 1$, $a'X_{j+1} = 0$ and $a'(X_{j+1} - D\bar{x}_{j+1}) = 0$, but it
is impossible for any weight vector $a$ to satisfy these conditions.\footnote{This
  follows because, under lack of overlap, the $j$th column of $D \circ
  \tilde{X}_{-1}$ equals $X_{j+1} - D\bar{x}_{j+1}$.} Hence,
no unbiased linear estimator exists, and the worst-case bias of any linear
estimator is unbounded. Appendix~\ref{appsec:imposs} gives the formal proof.

Adaptive CIs are also impossible. Such CIs would shrink with true VCATE while
maintaining coverage under a conservative \textit{a priori} bound. Corollary 2.1 of
\citet{armstrong2023biasaware} implies that any valid CI must have expected length that
reflects the conservative \textit{a priori} bound $C^2$, even when VCATE is much smaller
than $C^2$. Hence, the bound cannot be automated when constructing the CI.

Accordingly, $C$ cannot be set to infinity or selected from the data with the
goal of adapting to the true VCATE.
Section~\ref{sec:sensitivity} discusses sensitivity analysis with respect to $C$.

\section{Sensitivity analysis and applications} \label{sec:sensitivity}

Researchers often use the short regression under the implicit assumption that
treatment effect heterogeneity is limited. The proposed procedure assesses this
assumption by starting at $C = 0$, which imposes constant effects, and increasing
$C$ until the result becomes insignificant.\footnote{The sensitivity parameter is $C$, rather than $C^2$,
because it has the same units as the outcome.} Let $C^\ast$ denote this
breakdown value. Because VCATE is an interpretable dispersion measure,
researchers can compare $C^\ast$ with externally motivated reference values, as in
\citet{kline2013sensitivity}, \citet{masten2020inference}, and
\citet{li2021linear}. The exercise therefore reports how much departure from
constant effects can be accommodated before the inferential conclusion changes.
The R package provides plotting functionality.

\subsection{Staggered adoption}\label{sec:staggered-application}

The empirical application follows \citet{sun_estimating_2020}, which builds on
\citet{dobkin_finkelstein_kluender_notowidigdo_aer2018} to estimate the effect
of unexpected hospitalization on out-of-pocket medical spending in the Health
and Retirement Study (HRS). Each HRS wave spans about two calendar years. Using
a balanced five-period panel from wave 7 through wave 11, all individuals are
treated in the final period, so the ATT from waves
8 to 11 is not point-identified and the saturated long
specification in~\eqref{eq:saturated TWFE} cannot be estimated without further
restrictions. The ``static''
specification nonetheless gives a
statistically significant positive estimate.

\begin{figure}[H]
    \begin{center}
        \caption{\label{fig:oop_spend}Sensitivity Analysis for \citet{dobkin_finkelstein_kluender_notowidigdo_aer2018}}
\includegraphics[width=0.8\linewidth]{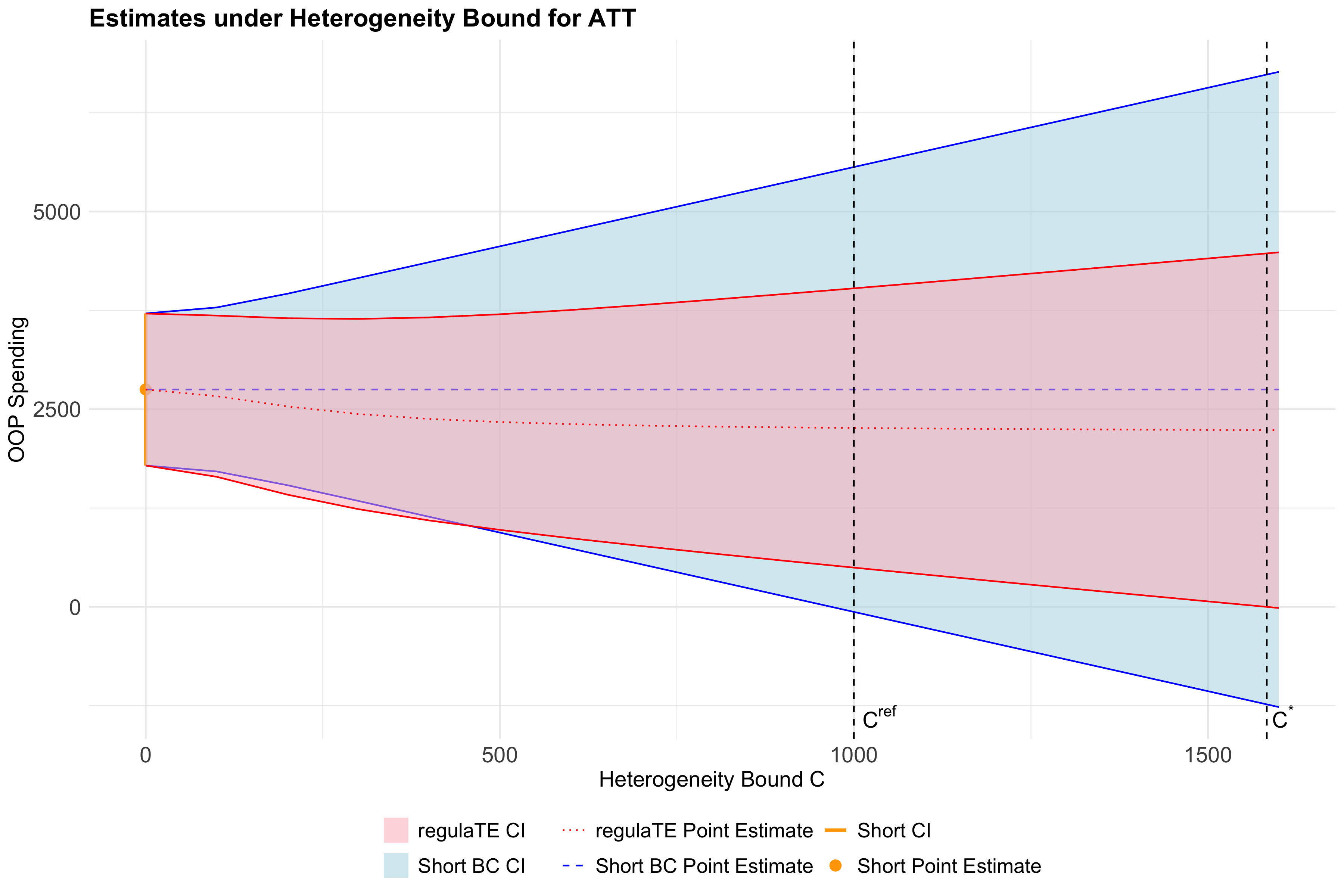}

        \end{center}
    \footnotesize{Notes: ``\texttt{regulaTE} Point Estimate'' refers to the \texttt{regulaTE} estimate optimized under each heterogeneity bound $C$ on the horizontal axis. ``Short Point Estimate'' and ``Short BC Point Estimate'' both refer to the short/static regression~\eqref{eq:static TWFE} estimate. ``\texttt{regulaTE} CI'' and ``Short BC CI'' refer to the heterogeneity-aware fixed-length confidence intervals based on the corresponding estimates, which are asymptotically valid at the 95\% level under each heterogeneity bound $C$.}
\end{figure}

Figure~\ref{fig:oop_spend} reports \texttt{regulaTE} CIs, with bias-corrected
short-regression CIs for comparison. Standard errors are clustered at the
individual level, as in the original analysis. The \texttt{regulaTE} CIs remain significant up to
$C^\ast = \$1{,}583$. The original analysis reports an out-of-pocket spending
increase of about \$3{,}000 in the first year after hospitalization and
\$1{,}000 by the third year, reflecting its focus on heterogeneity over time.
Treating these time-varying estimates as rough bounds, Popoviciu's
inequality gives $C^{\text{ref}} = \$1{,}000$. Because $C^\ast$ exceeds
$C^{\text{ref}}$, the \texttt{regulaTE} CI remains significant at the reference
value. The bias-corrected short-regression CIs widen more rapidly and reach zero
at a smaller value of $C$, indicating greater sensitivity when the
short-regression weights are held fixed. Thus, the positive ATT remains
statistically significant at the externally motivated reference
value under the \texttt{regulaTE} procedure.

\subsection{Unconfoundedness}
\citet{aizer2016} use administrative data to evaluate the long-run effects of
early 20th-century Mothers' Pension (MP) cash transfers on children's lifetime
outcomes. Their study
compares children of approved MP applicants to those whose approvals were
subsequently reversed, while controlling for observed characteristics, and
estimates the short regression
\citep[Equation (1)]{aizer2016}:
\[ 
\log(\textit{Age at Death})_{ifts} = \theta_0 + \theta_1 MP_f + \theta_2 \boldsymbol{X}_{if} + \theta_3\boldsymbol{Z}_{st} + \boldsymbol{\theta}_c + \boldsymbol{\theta}_t + \varepsilon_{if},
\]
where the outcome is log age at death for individual $i$ in family $f$, born in
year $t$ and living in county $c$ in state $s$, and $MP_f$ indicates MP receipt.
Controls include family and child characteristics $\boldsymbol X_{if}$,
county-level and state characteristics $\boldsymbol Z_{st}$, and county and
cohort fixed effects. The application focuses on child longevity, one of the
main outcomes in the original study.

The analysis assesses the ATT, defined as the average impact of MP receipt among
recipients. This is the relevant target for the program evaluation.
The short-regression estimate of the ATT is $1.82\%$ and is statistically
significant at the $5\%$ level. The long regression is infeasible because all
applicants received MP in eleven counties, which account for about 9\% of the
sample. The resulting collinearity among the MP receipt indicator, interaction
terms, and county fixed effects leaves the long regression undefined.

After reporting the ATT, the original paper examines subgroup effects by family
income, child's age, and urban residence. Some subgroup estimates are
statistically insignificant; the point estimates range between 1\% and 2\%
\citep[Table 5]{aizer2016}. The range of these estimates provides a reference for the amount
of heterogeneity considered below.

\begin{figure}[htbp!]   
    \begin{center}
    \caption{\label{fig:specification_3}Sensitivity Analysis for \citet{aizer2016}}
        \includegraphics[width=0.8\linewidth]{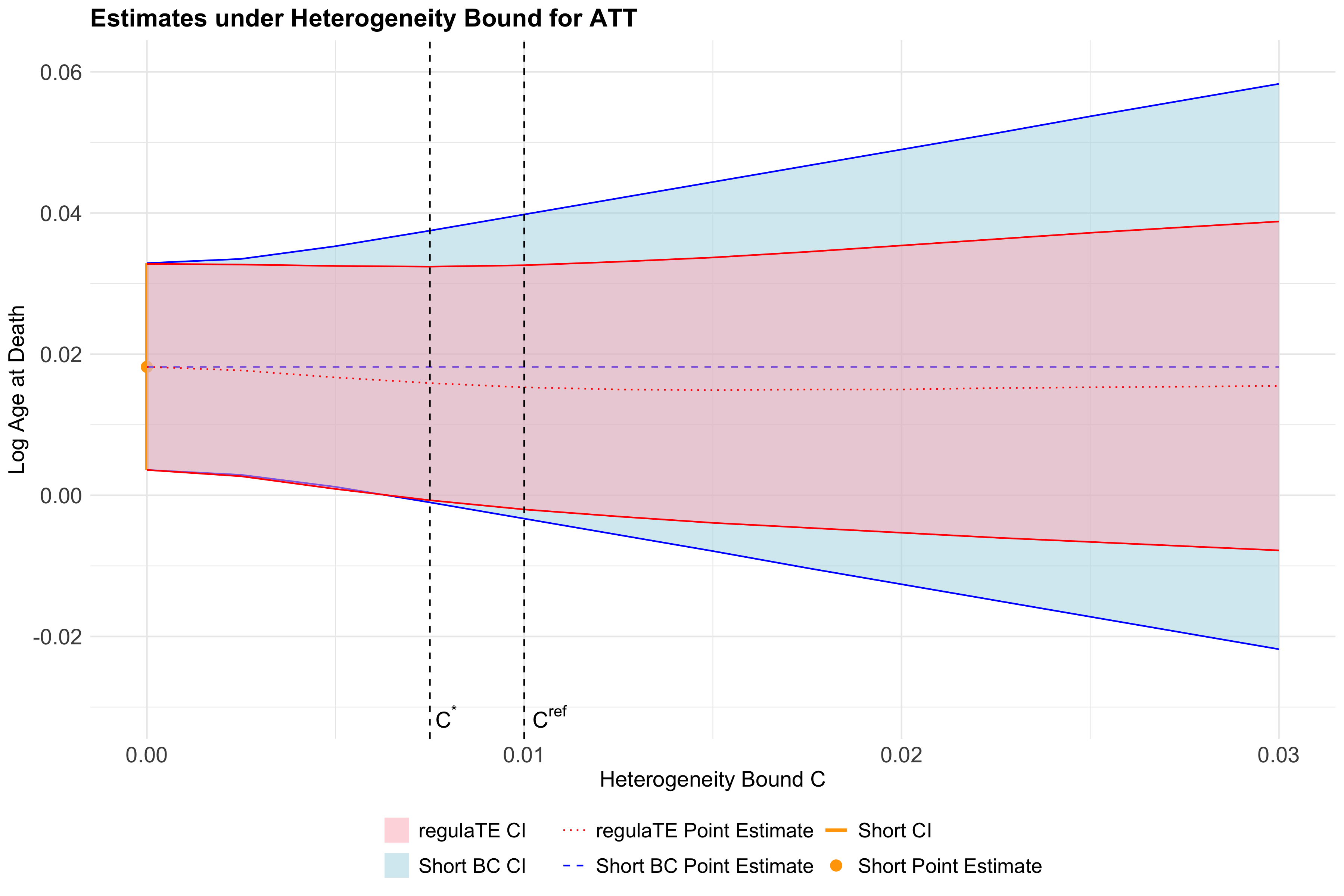}
    \end{center}
    \footnotesize{Notes: ``\texttt{regulaTE} Point Estimate'' refers to the \texttt{regulaTE} estimate optimized under each heterogeneity bound $C$ on the horizontal axis. ``Short Point Estimate'' and ``Short BC Point Estimate'' both refer to the short regression estimate. ``\texttt{regulaTE} CI'' and ``Short BC CI'' refer to the heterogeneity-aware fixed-length confidence intervals based on the corresponding estimates, which are asymptotically valid at the 95\% level under each heterogeneity bound $C$.}
\end{figure}

To relate the sensitivity analysis to the subgroup results,
Figure~\ref{fig:specification_3} reports the \texttt{regulaTE} CI and the
bias-corrected short-regression CI. Standard errors are clustered at the county
level, as in the original analysis. As $C$ increases, the \texttt{regulaTE} CI
expands to account for worst-case bias but remains shorter than the
bias-corrected short-regression CI. The breakdown value is 0.75\%
($C^\ast=0.0075$). Taking 0\% and 2\% as conservative bounds on treatment
effects, consistent with the subgroup estimates, Popoviciu's inequality gives
1\% as an upper bound on the standard deviation. We therefore use
$C^{\text{ref}} = 1\%$. The breakdown value is
therefore 0.25 percentage points below $C^{\text{ref}}$: statistical
significance is lost before the reference amount of heterogeneity is reached.

\section{Conclusion}\label{sec:conclusion}
Many standard empirical specifications restrict treatment effects to be
constant. This restriction can shorten confidence intervals but can also induce
bias under treatment effect heterogeneity. This paper analyzes sensitivity to
the restriction by varying a bound on the target-weighted variance of treatment
effects, thereby making the efficiency-robustness trade-off explicit. We conduct this sensitivity analysis using our proposed generalized ridge estimator, \texttt{regulaTE}. At $C=0$,
the analysis corresponds to the constant-effects restriction
embedded in the short regression. As $C$ increases, the confidence intervals
allow larger departures from constant effects, including cases in which limited
overlap makes the fully interacted long regression unstable or unidentified. In both empirical applications, \texttt{regulaTE} yields shorter confidence intervals than the bias-corrected short
regression, thereby enabling informative sensitivity analysis over a range of heterogeneity bounds.

Future work could consider other restrictions on treatment effect heterogeneity,
such as absolute bounds on deviations from the average effect when prior
knowledge supports them. Another extension is to allow multiple treatments or
multiple sources of treatment effect heterogeneity, as in
\citet{goldsmithpinkham2024contamination} and \citet{sun_estimating_2020}.

\newpage
\appendix
\setcounter{equation}{0}
\renewcommand{\theequation}{\thesection.\arabic{equation}}
\setcounter{theorem}{0}
\renewcommand{\thetheorem}{\thesection.\arabic{theorem}}
\setcounter{lemma}{0}
\renewcommand{\thelemma}{\thesection.\arabic{lemma}}
\setcounter{assumption}{0}
\renewcommand{\theassumption}{\thesection.\arabic{assumption}}
\setcounter{claim}{0}
\renewcommand{\theclaim}{\thesection.\arabic{claim}}

\section{Proofs of Main Results}
This appendix follows the order of the main text. Section~A.1 proves the
auxiliary results and Theorem~\ref{thm:ridge frontier}; the Online Appendix
gives the longer discrete-covariate derivations and no-overlap limit.
Section~A.2 proves Theorem~\ref{general_se_thm}, and Section~A.3 proves
Proposition~\ref{thm:impossibility}.

\subsection{Proofs for Section~\ref{sec:bias-variance}: Lemmas~\ref{claim:short bias} and~\ref{lem:outcome ridge}, and Theorem~\ref{thm:ridge frontier}}

The results in this subsection retain the general distinction between the
control matrix $W$ and the effect-modifier matrix $X$.

\begin{lemma}\label{claim:short bias}

The short regression estimator is $\hat{\beta}_{\text{short}}=a_{s}'Y$, where
$a_{s}=(D'H_{W}D)^{-1}H_{W}D$. Under
model~\eqref{eq: reg model fs} and the bound $\delta'V_x\delta\leq C^2$, its
worst-case bias is
\[
\maxbias_{a_{s},C}=C\cdot\sqrt{a_{s}'(D\circ\tilde{X}_{-1})V_{x}^{-1}(D\circ\tilde{X}_{-1})'a_{s}}.
\]
\end{lemma}
\begin{proof}[Proof for Lemma~\ref{claim:short bias}]
The expression for $a_{s}$ follows from the Frisch--Waugh--Lovell theorem. The worst-case bias is 
\begin{align}
\begin{aligned}\maxbias_{a_{s},C}:= & \sup_{(\beta,\gamma',\delta')'\in\Theta_{C}}\E[a_{s}'Y]-\beta\\
= & \sup_{(\beta,\gamma',\delta')'\in\Theta_{C}}a_{s}'D\beta+a_{s}'(D\circ\tilde{X}_{-1})\delta+a_{s}'W\gamma-\beta\\
= & \sup_{\delta'V_{x}\delta\leq C^{2}}a_{s}'(D\circ\tilde{X}_{-1})\delta,
\end{aligned}\label{eq:short bias}
\end{align}
where the last equality holds because $a_{s}'D=1$ and $a_{s}'W=0$.
If $(D\circ\tilde{X}_{-1})'a_s=0$, the result is immediate. Otherwise, the
quadratic constraint binds, and the Karush--Kuhn--Tucker condition gives
\[
\delta=V_{x}^{-1}(D\circ\tilde{X}_{-1})'a_{s}\sqrt{\frac{C^{2}}{a_{s}'(D\circ\tilde{X}_{-1})V_{x}^{-1}(D\circ\tilde{X}_{-1})'a_{s}}}.
\]

Substitution gives the expression for the worst-case
bias.
\end{proof}

\Needspace{10\baselineskip}
\begin{lemma}[Two-step representation of generalized ridge regression]\label{lem:outcome ridge}
For a matrix of nuisance regressors $R$, consider the generalized ridge estimator
\[
\min_{\beta,\eta}\norm{Y-D\beta-R\eta}^{2}+\eta'A\eta 
\]
where $A$ is positive semidefinite, $\left(R'R+A\right)$ is invertible, and
\[
D'\left[I-R(R'R+A)^{-1}R'\right]D>0
\]
is assumed. The solution $\beta^\ast$ can be written as
\[
\beta^\ast=\frac{\left(D -R \tilde\eta\right)'Y }{ \left(D -R\tilde\eta\right)'D},
\]
where $\tilde\eta$ solves
\[
\min_{\eta}\norm{D -R \eta}^{2} +\eta'A\eta.
\]

\end{lemma}

\begin{proof}[Proof for Lemma~\ref{lem:outcome ridge}]
Let $(\beta^{\ast},\eta^{\ast})$ denote the solution to the generalized ridge regression. The first-order
condition for $\eta$ gives $\eta^{\ast}=\left(R'R+A\right)^{-1}R'\left(Y-D\beta^{\ast}\right)$.
 Substituting this into the first-order condition for $\beta$ gives
\begin{equation*}
    D'\left(Y -R \left(R'R+A\right)^{-1}R'Y\right) =D'\left(D -R \left(R'R+A\right)^{-1}R'D\right)\beta^{\ast}.
\end{equation*}
The left-hand side can be written as
\begin{equation*}
 D'Y - D'R\left(R'R+A\right)^{-1}R'Y = \left(D - R\left(R'R+A\right)^{-1}R'D\right)'Y.
\end{equation*}
Also, $\left(R'R+A\right)^{-1}R'D$
solves 
\[
\min_{\eta}\norm{D -R \eta}^{2} +\eta'A\eta.
\]
The additional condition makes the denominator
$\left(D-R\tilde\eta\right)'D$ strictly positive. Combining these expressions
yields the result.
\end{proof}

\begin{proof}[Proof of Theorem \ref{thm:ridge frontier}]

Let $(\pi_{1,\lambda},\pi_{2,\lambda})$ be the ridge coefficients solving
\[
\min_{\pi_{1},\pi_{2}}\;\norm{D-W\pi_{1}-(D\circ\tilde{X}_{-1})\pi_{2}}^{2}/n+\lambda\,\pi_{2}'V_{x}\pi_{2}.
\]
Following the derivation of~\eqref{eq:opt_sol}, the associated weights are based
on
\[
\tilde{D}_{\lambda}:=D-W\pi_{1,\lambda}-(D\circ\tilde{X}_{-1})\pi_{2,\lambda},\qquad a_{\lambda}:=\frac{\tilde{D}_{\lambda}}{\tilde{D}_{\lambda}'D}.
\]

The first-order conditions with respect to $\pi_{1,\lambda}$ and $\pi_{2,\lambda}$
are 
\begin{equation}
W'\tilde{D}_{\lambda}=0,\ (D\circ\tilde{X}_{-1})'\tilde{D}_{\lambda}/n=\lambda V_{x}\pi_{2,\lambda}.\label{eq:foc-pi2}
\end{equation}

Since $W'a_{\lambda}=0$, the derivation for the worst-case
bias in~\eqref{eq:short bias} still applies and the term within the square root is equal to
\[
a_{\lambda}'(D\circ\tilde{X}_{-1})V_{x}^{-1}(D\circ\tilde{X}_{-1})'a_{\lambda}=\frac{\tilde{D}_{\lambda}'(D\circ\tilde{X}_{-1})V_{x}^{-1}(D\circ\tilde{X}_{-1})'\tilde{D}_{\lambda}}{(\tilde{D}_{\lambda}'D)^{2}}.
\]
Applying \eqref{eq:foc-pi2},
\[
\tilde{D}_{\lambda}'(D\circ\tilde{X}_{-1})V_{x}^{-1}(D\circ\tilde{X}_{-1})'\tilde{D}_{\lambda}=n^{2}\lambda^{2}\,\pi_{2,\lambda}'V_{x}\pi_{2,\lambda},
\]
 and $\tilde{D}_{\lambda}'D>0$ because 
\begin{align*}
\tilde{D}_{\lambda}'D & =\tilde{D}_{\lambda}'\left(\tilde{D}_{\lambda}+W\pi_{1,\lambda}+(D\circ\tilde{X}_{-1})\pi_{2,\lambda}\right)=\tilde{D}_{\lambda}'\tilde{D}_{\lambda}+\tilde{D}_{\lambda}'(D\circ\tilde{X}_{-1})\pi_{2,\lambda}\\
 & =\tilde{D}_{\lambda}'\tilde{D}_{\lambda}+n\lambda\,\pi_{2,\lambda}'V_{x}\pi_{2,\lambda}>0.
\end{align*}
It follows that
\[
\sqrt{a_{\lambda}'(D\circ\tilde{X}_{-1})V_{x}^{-1}
(D\circ\tilde{X}_{-1})'a_{\lambda}}
=\frac{n\lambda\sqrt{\pi_{2,\lambda}'V_{x}\pi_{2,\lambda}}}
{\tilde{D}_{\lambda}'D}.
\]

If $\pi_{2,\lambda}=0$, then $\tilde D_\lambda=H_WD$ and
$a_\lambda=a_s$, while~\eqref{eq:foc-pi2} gives
$(D\circ\tilde X_{-1})'a_\lambda=0$. Thus, $B=0$. For every feasible $a$,
$1=a'H_WD\leq\norm{a}\norm{H_WD}$, with equality at $a_\lambda$; hence,
$a_\lambda$ solves~\eqref{eq:bias-var}. We may therefore assume
$\pi_{2,\lambda}'V_x\pi_{2,\lambda}>0$.

Rewrite this expression as
\begin{equation}
\frac{n\lambda\sqrt{\pi_{2,\lambda}'V_{x}\pi_{2,\lambda}}}{\tilde{D}_{\lambda}'D}=\frac{n\lambda\pi_{2,\lambda}'V_{x}\pi_{2,\lambda}}{\tilde{D}_{\lambda}'D\sqrt{\pi_{2,\lambda}'V_{x}\pi_{2,\lambda}}}.\label{eq:quadform}
\end{equation}

Equation~\eqref{eq:foc-pi2} also gives
\[
\begin{aligned}
n\lambda\pi_{2,\lambda}'V_{x}\pi_{2,\lambda}
&=\tilde{D}_{\lambda}'(D\circ\tilde{X}_{-1})\pi_{2,\lambda}\\
&=\tilde{D}_{\lambda}'(W\pi_{1,\lambda}
  +(D\circ\tilde{X}_{-1})\pi_{2,\lambda}).
\end{aligned}
\]

Substituting this identity into \eqref{eq:quadform}
and using the definition of $a_{\lambda}$ gives 
\[
\sqrt{a_{\lambda}'(D\circ\tilde{X}_{-1})V_{x}^{-1}(D\circ\tilde{X}_{-1})'a_{\lambda}}=\frac{a_{\lambda}'(W\pi_{1,\lambda}+(D\circ\tilde{X}_{-1})\pi_{2,\lambda})}{\sqrt{\pi_{2,\lambda}'V_{x}\pi_{2,\lambda}}},
\]
which implies that the worst-case bias equals
 \begin{equation}\label{eq:wc-bias}
\maxbias_{a_{\lambda}, C} = C \frac{a_{\lambda}'(W\pi_{1, \lambda} + (D \circ
  \tilde{X}_{-1}) \pi_{2, \lambda})}{\sqrt{\pi_{2, \lambda}'V_{x}\pi_{2, \lambda}}}.
\end{equation} 

The problem satisfies the conditions of Theorem~2.1 in
\citet{armstrong2023biasaware}. In their notation, the estimand is the linear functional
    $L(\theta)=\beta$, the observed signal is
    $f(\theta)=D\beta+W\gamma+(D\circ\tilde X_{-1})\delta$, the nuisance
    parameters are $(\gamma',\delta')'$, and the penalty is
    $\mathrm{Pen}((\gamma',\delta')')=\sqrt{\delta'V_x\delta}$, which is a
    seminorm. The constraint set $\Theta_C$ corresponds to
    $\{\theta:\mathrm{Pen}((\gamma',\delta')')\le C\}$ with $\beta$
    unrestricted. Hence, that theorem implies that the weights $a_\lambda$, which
    arise from the penalized regression~\eqref{eq:gen_ridge} with penalty
    parameter $\lambda$, solve the bias-constrained variance-minimization
    problem~\eqref{eq:bias-var} with $B=\maxbias_{a_\lambda,C}$ as
    in~\eqref{eq:wc-bias}. 
 \end{proof} 

\subsubsection{Additional proofs for discrete covariates}

The Online Appendix proves the discrete-covariate representation, closed-form
weights, continuity result for vanishing ridge penalties, and no-overlap
trimming limit used in Section~\ref{sec:bias-variance}.

\subsection{Proof of Theorem~\ref{general_se_thm} for Section~\ref{sec:implementation}}\label{sec:feasible CI primitive condition}
We adapt Appendix B.2 of \citet{armstrong2023biasaware}. The distribution $Q$ of
$\varepsilon$ may be unknown and non-Gaussian; only independence across $i$ is
imposed. For independent clusters, the same argument applies under corresponding
Lindeberg and variance consistency conditions.

Let $\mathcal{Q}_n$ denote the class of possible error distributions $Q$. Let
$P_{\theta, Q}$ and $E_{\theta, Q}$ denote
probability and expectation when $Y$ is drawn according to $Q\in \mathcal{Q}_n$
and $\theta\in\Theta_C$, and let $P_Q$ and $E_Q$ denote expressions
that depend on $Q$ only and not on $\theta$. Let
$M = [D,\, W,\, D \circ \tilde{X}_{-1}]$ denote the full design matrix.

Consider the estimator introduced in Section~\ref{sec:implementation}. Let
\[
V_{Q}
=\Var_{Q}(\hat{\beta}_{\lambda^*_{C}})
=\sum_{i=1}^{n}a_{\lambda^*_{C},i}^2E_{Q}\varepsilon_i^2
\]
denote the variance of the estimator.

\begin{theorem}\label{general_se_thm}
  Suppose that, for some $\eta>0$,
  $\eta\le E_Q[\varepsilon_i^2]$ and
  $E_Q[\abs{\varepsilon_i}^{2+\eta}] \le 1/\eta$ for all $i$ and all
  $Q\in\mathcal{Q}_n$. Suppose also that the following conditions hold for some
  sequence $c_n$ with $c_n=\mathcal{O}(\sqrt{n})$:
  \begin{enumerate}[label=({\roman*})]
  \item $\max\left\{\sqrt{n}c_n,1\right\}\cdot \operatorname{Lind}(a_{\lambda^*_{C}}) \to
    0$; and
  \item $\inf_{\theta\in\Theta_C, Q\in\mathcal{Q}_n} P_{\theta, Q}
    (\norm{M(\hat\theta^{\mathrm{init}}-\theta)}_2 \le c_n)\to 1$.
  \end{enumerate}
  Then, for any $\zeta>0$,
  $\inf_{\theta\in\Theta_C, Q\in\mathcal{Q}_n} P_{\theta,Q}\left(\abs{(\hat{V}_{\lambda^*_{C},\text{rob}}-V_Q)/V_Q} <
    \zeta \right) \to 1$. Furthermore,
  \begin{equation}\label{eq:feasible_flci_coverage}
    \liminf_n\inf_{\theta\in\Theta_C, Q\in\mathcal{Q}_n} P_{\theta,Q}\left(
      \beta\in \left\{\hat\beta_{\lambda^*_{C}}\pm
        \cv_\alpha(\maxbias_{a_{\lambda^*_{C}},C}/\sqrt{\hat{V}_{\lambda^*_{C},\text{rob}}})\cdot
        \sqrt{\hat{V}_{\lambda^*_{C},\text{rob}}} \right\} \right) \ge 1-\alpha.
  \end{equation}
\end{theorem}
The convergence rate $c_n$ depends on whether
$\hat\theta^{\mathrm{init}}$ is estimated by the long regression or by
cross-validated generalized ridge regression.

\begin{proof}
The argument adapts Appendix~B.2 of \citet{armstrong2023biasaware}. The
estimator $\hat\beta_{\lambda^*_C}=a_{\lambda^*_C}'Y$ is a linear estimator
with nonrandom weights applied to independent errors. Under condition~(i),
the Lindeberg condition holds for the triangular array
$\{a_{\lambda^*_C,i}\varepsilon_i\}_{i=1}^n$, yielding
$(\hat\beta_{\lambda^*_C}-\E[\hat\beta_{\lambda^*_C}])/V_Q^{1/2}
\xrightarrow{d}N(0,1)$. For the variance estimator, writing
$\hat\varepsilon_{\mathrm{init},i}=\varepsilon_i +
M_i'(\theta-\hat\theta^{\mathrm{init}})$, condition~(ii) ensures the
cross-terms are asymptotically negligible, so that
$|\hat{V}_{\lambda^*_C,\text{rob}}-V_Q|/V_Q\xrightarrow{p}0$. The coverage
result~\eqref{eq:feasible_flci_coverage} then follows from the bias bound
$|\E[\hat\beta_{\lambda^*_C}]-\beta|
\le\maxbias_{a_{\lambda^*_C},C}$ and Slutsky's theorem.
\end{proof}

\subsection{Proof of Proposition~\ref{thm:impossibility} for Section~\ref{sec:nec_bd_vcate}}
\label{appsec:imposs}
As in Section~\ref{sec:nec_bd_vcate}, this result specializes to the
unconfoundedness design with $W=X$.
Any CI $\mathcal{C}$ with uniformly valid coverage over the unrestricted
parameter space has unbounded worst-case expected length:
$\sup_{(\beta, \gamma', \delta')\in\mathbb{R}^{2 + 2k}}
\E_{(\beta, \gamma', \delta')} \mu(\mathcal{C}) = \infty$, where $\mu$ denotes
Lebesgue measure. This follows from the worst-case CI length bound of
\citet{low1997nonparametric} for the linear functions considered here.

\begin{claim}\label{thm:impossibility}
  Let $\mathcal{C}$ be a CI for $\beta$ with nominal coverage probability
  $1-\alpha$ (i.e.,
  $\inf_{(\beta, \gamma', \delta') \in \mathbb{R}^{2 + 2k}} P(\beta \in
  \mathcal{C}) \geq 1-\alpha$). If there is no overlap, then
  $\sup_{(\beta, \gamma', \delta') \in \mathbb{R}^{2 + 2k}} \E_{(\beta, \gamma',
    \delta')} \mu(\mathcal{C}) = \infty$.
\end{claim}

\begin{remark}
  The proposition applies to any CI $\mathcal{C}$ without restricting
  it to be a fixed-length CI based on affine estimators. Hence, the result
  establishes that some parameter-space restriction is necessary for informative
  CIs when overlap fails. For fixed-length CIs, including the CIs studied here,
  the result implies $\mu(\mathcal{C}) = \infty$: no nontrivial CI of the usual
  ``linear estimator $\pm$ critical value'' form has correct coverage.
\end{remark}

\begin{proof}
  By \citet{low1997nonparametric}, the worst-case length is lower bounded by
  $(1-\alpha- \epsilon/4) \omega(\epsilon)$, where $\omega(\epsilon)$
  is the \emph{modulus of continuity} (see, e.g., \citealp{donoho1994statistical}
  or \citealp{armstrong_optimal_2018}) defined as
  \begin{equation*}
    \omega(\epsilon) := \sup_{(\beta, \gamma', \delta')' \in \mathbb{R}^{2 + 2k}} 2\beta
    \quad \text{s.t.} \quad \norm{D\beta +
  X\gamma + (D \circ \tilde{X}_{-1})\delta}^{2} \leq \frac{\epsilon^{2}}{4}.
\end{equation*}
It suffices to show that $\omega(\epsilon) = \infty$. Suppose to the
contrary that $\omega(\epsilon) < \infty$. Then, there exists $\theta^{\ast}=
(\beta^{\ast},
\gamma^{\ast}{}', \delta^{\ast}{}')'$ such that $\theta^{\ast}$ satisfies the
constraint and $\omega(\epsilon) = 2\beta^{\ast}$. It must be the
case that $\beta^{\ast} > 0$. Since there is no overlap, there exists some
$j$ such that, without loss of generality, the binary $X_{j+1} \leq D$,
where $X_{j+1}$ denotes the ($j+1$)th column of $X$, and $\leq$ is interpreted
elementwise. That is, all individuals $i$ with $X_{i,j+1} = 1$ are treated.

Consider the $j$th column of $D \circ \tilde{X}_{-1}$, which is by definition $D \circ
(X_{j+1} - \bar{x}_{j+1}\mathbf{1})$. The identities $D \circ
(\bar{x}_{j+1}\mathbf{1}) = D \bar{x}_{j+1}$ and $D \circ
X_{j+1} = X_{j+1}$ hold under lack of overlap. For any constant $c > 0$, consider $\theta^{\ast}_{c} =
(\beta^{\ast} + c \bar{x}_{j+1}, \gamma^{\ast} - c e_{j+1}, \delta^{\ast} + ce_{j})$. This
vector satisfies the constraint because
\begin{align*}
&D(\beta^\ast + c \bar{x}_{j+1}) + X(\gamma^{\ast} - c e_{j+1}) + (D \circ \tilde{X}_{-1})(
  \delta^{\ast} + ce_{j}) \\
  = & D\beta^\ast + X\gamma^{\ast} + (D \circ \tilde{X}_{-1})
  \delta^{\ast} + (c \bar{x}_{j+1})D - c X_{j+1} + c (D \circ
      (X_{j+1} - \bar{x}_{j+1}\mathbf{1})) \\
  = & D\beta^\ast + X\gamma^{\ast} + (D \circ \tilde{X}_{-1})
  \delta^{\ast} + (c \bar{x}_{j+1})D - c X_{j+1} + c (X_{j+1} - \bar{x}_{j+1}D)
  \\
  = &  D\beta^\ast + X\gamma^{\ast} + (D \circ \tilde{X}_{-1})
  \delta^{\ast}.
\end{align*}
Since $\beta^{\ast} + c \bar{x}_{j+1} > \beta^{\ast}$, this is a
contradiction. Hence, it must be the case that $\omega(\epsilon) = \infty$.
\end{proof}

\bibliographystyle{chicago}
\bibliography{./ref.bib}

\begin{thebibliography}{}

\bibitem[\protect\citeauthoryear{Aizer, Eli, Ferrie, and Lleras-Muney}{Aizer
  et~al.}{2016}]{aizer2016}
Aizer, A., S.~Eli, J.~Ferrie, and A.~Lleras-Muney (2016).
\newblock The long-run impact of cash transfers to poor families.
\newblock {\em American Economic Review\/}~{\em 106\/}(4), 935–71.

\bibitem[\protect\citeauthoryear{Angrist}{Angrist}{1998}]{angrist1998}
Angrist, J.~D. (1998).
\newblock Estimating the labor market impact of voluntary military service
  using social security data on military applicants.
\newblock {\em Econometrica\/}~{\em 66\/}(2), 249--288.

\bibitem[\protect\citeauthoryear{Angrist and Pischke}{Angrist and
  Pischke}{2009}]{angrist2009mostly}
Angrist, J.~D. and J.-S. Pischke (2009).
\newblock {\em Mostly harmless econometrics: An empiricist's companion}.
\newblock Princeton university press.

\bibitem[\protect\citeauthoryear{Armstrong and Koles\'{a}r}{Armstrong and
  Koles\'{a}r}{2018}]{armstrong_optimal_2018}
Armstrong, T.~B. and M.~Koles\'{a}r (2018).
\newblock Optimal inference in a class of regression models.
\newblock {\em Econometrica\/}~{\em 86\/}(2), 655--683.

\bibitem[\protect\citeauthoryear{Armstrong and Koles{\'a}r}{Armstrong and
  Koles{\'a}r}{2021}]{armstrong2021finite}
Armstrong, T.~B. and M.~Koles{\'a}r (2021).
\newblock Finite-sample optimal estimation and inference on average treatment
  effects under unconfoundedness.
\newblock {\em Econometrica\/}~{\em 89\/}(3), 1141--1177.

\bibitem[\protect\citeauthoryear{Armstrong, Kolesár, and Kwon}{Armstrong
  et~al.}{2023}]{armstrong2023biasaware}
Armstrong, T.~B., M.~Kolesár, and S.~Kwon (2023).
\newblock Bias-aware inference in regularized regression models.
\newblock {\em arXiv:2012.14823 [econ.EM]\/}.

\bibitem[\protect\citeauthoryear{Athey, Imbens, and Wager}{Athey
  et~al.}{2018}]{athey_approximate_2018}
Athey, S., G.~W. Imbens, and S.~Wager (2018).
\newblock Approximate residual balancing: debiased inference of average
  treatment effects in high dimensions.
\newblock {\em Journal of the Royal Statistical Society: Series B (Statistical
  Methodology)\/}~{\em 80\/}(4), 597--623.

\bibitem[\protect\citeauthoryear{Bobonis, C{\'a}mara~Fuertes, and
  Schwabe}{Bobonis et~al.}{2016}]{bobonis2016monitoring}
Bobonis, G.~J., L.~R. C{\'a}mara~Fuertes, and R.~Schwabe (2016).
\newblock Monitoring corruptible politicians.
\newblock {\em American Economic Review\/}~{\em 106\/}(8), 2371--2405.

\bibitem[\protect\citeauthoryear{Callaway and Sant’Anna}{Callaway and
  Sant’Anna}{2021}]{callaway_santanna_ssrn2018}
Callaway, B. and P.~H. Sant’Anna (2021).
\newblock Difference-in-differences with multiple time periods.
\newblock {\em Journal of econometrics\/}~{\em 225\/}(2), 200--230.

\bibitem[\protect\citeauthoryear{Chernozhukov, Deaner, Gao, Hausman, and
  Newey}{Chernozhukov et~al.}{2025}]{chernozhukov2025fisher}
Chernozhukov, V., B.~Deaner, Y.~Gao, J.~A. Hausman, and W.~Newey (2025).
\newblock Linear estimation of structural and causal effects for nonseparable
  panel data.
\newblock Technical report, National Bureau of Economic Research.

\bibitem[\protect\citeauthoryear{Crump, Hotz, Imbens, and Mitnik}{Crump
  et~al.}{2009}]{crump_dealing_2009}
Crump, R.~K., V.~J. Hotz, G.~W. Imbens, and O.~A. Mitnik (2009).
\newblock Dealing with limited overlap in estimation of average treatment
  effects.
\newblock {\em Biometrika\/}~{\em 96\/}(1), 187--199.

\bibitem[\protect\citeauthoryear{de~Chaisemartin}{de~Chaisemartin}{2024}]{dechaisemartin2021tradingoff}
de~Chaisemartin, C. (2024).
\newblock Trading-off bias and variance in stratified experiments and in
  staggered adoption designs, under a boundedness condition on the magnitude of
  the treatment effect.
\newblock {\em arXiv:2105.08766 [econ.EM]\/}.

\bibitem[\protect\citeauthoryear{de~Chaisemartin and Deeb}{de~Chaisemartin and
  Deeb}{2024}]{dechaisemartin2024estimatingtreatmenteffectheterogeneitysites}
de~Chaisemartin, C. and A.~Deeb (2024).
\newblock Estimating treatment-effect heterogeneity across sites, in multi-site
  randomized experiments with few units per site.
\newblock {\em arXiv:2405.17254 [econ.EM]\/}.

\bibitem[\protect\citeauthoryear{de~Chaisemartin and
  D'Haultf{\oe}uille}{de~Chaisemartin and D'Haultf{\oe}uille}{2020}]{double_fe}
de~Chaisemartin, C. and X.~D'Haultf{\oe}uille (2020).
\newblock Two-way fixed effects estimators with heterogeneous treatment
  effects.
\newblock {\em American Economic Review\/}~{\em 110\/}(9), 2964--96.

\bibitem[\protect\citeauthoryear{Dobkin, Finkelstein, Kluender, and
  Notowidigdo}{Dobkin
  et~al.}{2018}]{dobkin_finkelstein_kluender_notowidigdo_aer2018}
Dobkin, C., A.~Finkelstein, R.~Kluender, and M.~J. Notowidigdo (2018).
\newblock The economic consequences of hospital admissions.
\newblock {\em American Economic Review\/}~{\em 108\/}(2), 308--52.

\bibitem[\protect\citeauthoryear{Donoho}{Donoho}{1994}]{donoho1994statistical}
Donoho, D.~L. (1994).
\newblock Statistical estimation and optimal recovery.
\newblock {\em The Annals of Statistics\/}~{\em 22\/}(1), 238--270.

\bibitem[\protect\citeauthoryear{Favara and Imbs}{Favara and
  Imbs}{2015}]{favara2015credit}
Favara, G. and J.~Imbs (2015).
\newblock Credit supply and the price of housing.
\newblock {\em American economic review\/}~{\em 105\/}(3), 958--992.

\bibitem[\protect\citeauthoryear{Gibbons, Serrato, and Urbancic}{Gibbons
  et~al.}{2019}]{gibbons_serrato_urbancic_jem2018}
Gibbons, C.~E., J.~C.~S. Serrato, and M.~B. Urbancic (2019).
\newblock Broken or fixed effects?
\newblock {\em Journal of Econometric Methods\/}~{\em 8\/}(1), 1--12.

\bibitem[\protect\citeauthoryear{Goldsmith-Pinkham, Hull, and
  Koles{\'a}r}{Goldsmith-Pinkham
  et~al.}{2024}]{goldsmithpinkham2024contamination}
Goldsmith-Pinkham, P., P.~Hull, and M.~Koles{\'a}r (2024).
\newblock Contamination bias in linear regressions.
\newblock {\em American Economic Review\/}~{\em 114\/}(12), 4015--4051.

\bibitem[\protect\citeauthoryear{Goodman-Bacon}{Goodman-Bacon}{2021}]{goodman2021difference}
Goodman-Bacon, A. (2021).
\newblock Difference-in-differences with variation in treatment timing.
\newblock {\em Journal of Econometrics\/}~{\em 225\/}(2), 254--277.

\bibitem[\protect\citeauthoryear{Heiler and Kazak}{Heiler and
  Kazak}{2021}]{heiler2021valid}
Heiler, P. and E.~Kazak (2021).
\newblock Valid inference for treatment effect parameters under irregular
  identification and many extreme propensity scores.
\newblock {\em Journal of Econometrics\/}~{\em 222\/}(2), 1083--1108.

\bibitem[\protect\citeauthoryear{Imbens and Wooldridge}{Imbens and
  Wooldridge}{2009}]{imbens2009recent}
Imbens, G.~W. and J.~M. Wooldridge (2009).
\newblock Recent developments in the econometrics of program evaluation.
\newblock {\em Journal of economic literature\/}~{\em 47\/}(1), 5--86.

\bibitem[\protect\citeauthoryear{Kline, Saggio, and S{\o}lvsten}{Kline
  et~al.}{2020}]{kline2020leave}
Kline, P., R.~Saggio, and M.~S{\o}lvsten (2020).
\newblock Leave-out estimation of variance components.
\newblock {\em Econometrica\/}~{\em 88\/}(5), 1859--1898.

\bibitem[\protect\citeauthoryear{Kline and Santos}{Kline and
  Santos}{2013}]{kline2013sensitivity}
Kline, P. and A.~Santos (2013).
\newblock Sensitivity to missing data assumptions: Theory and an evaluation of
  the us wage structure.
\newblock {\em Quantitative Economics\/}~{\em 4\/}(2), 231--267.

\bibitem[\protect\citeauthoryear{Lechner}{Lechner}{2008}]{lechner2008note}
Lechner, M. (2008).
\newblock A note on the common support problem in applied evaluation studies.
\newblock {\em Annales d'{\'E}conomie et de Statistique\/}, 217--235.

\bibitem[\protect\citeauthoryear{Lee and Weidner}{Lee and
  Weidner}{2026}]{lee2021bounding}
Lee, S. and M.~Weidner (2026).
\newblock Bounding treatment effects by pooling limited information across
  observations.
\newblock {\em Journal of Econometrics\/}~{\em 256}, 106254.

\bibitem[\protect\citeauthoryear{Levy, van~der Laan, Hubbard, and
  Pirracchio}{Levy et~al.}{2021}]{levy2021fundamental}
Levy, J., M.~van~der Laan, A.~Hubbard, and R.~Pirracchio (2021).
\newblock A fundamental measure of treatment effect heterogeneity.
\newblock {\em Journal of Causal Inference\/}~{\em 9\/}(1), 83--108.

\bibitem[\protect\citeauthoryear{Li and M{\"u}ller}{Li and
  M{\"u}ller}{2021}]{li2021linear}
Li, C. and U.~K. M{\"u}ller (2021).
\newblock Linear regression with many controls of limited explanatory power.
\newblock {\em Quantitative Economics\/}~{\em 12\/}(2), 405--442.

\bibitem[\protect\citeauthoryear{Li}{Li}{1982}]{li1982minimaxity}
Li, K.-C. (1982).
\newblock Minimaxity of the method of regularization of stochastic processes.
\newblock {\em The Annals of Statistics\/}~{\em 10\/}(3), 937--942.

\bibitem[\protect\citeauthoryear{Low}{Low}{1997}]{low1997nonparametric}
Low, M.~G. (1997).
\newblock On nonparametric confidence intervals.
\newblock {\em The Annals of Statistics\/}~{\em 25\/}(6), 2547--2554.

\bibitem[\protect\citeauthoryear{Ma and Wang}{Ma and Wang}{2020}]{ma2020robust}
Ma, X. and J.~Wang (2020).
\newblock Robust inference using inverse probability weighting.
\newblock {\em Journal of the American Statistical Association\/}~{\em
  115\/}(532), 1851--1860.

\bibitem[\protect\citeauthoryear{Manski and Pepper}{Manski and
  Pepper}{2018}]{manski_how_2018}
Manski, C.~F. and J.~V. Pepper (2018).
\newblock {How Do Right-to-Carry Laws Affect Crime Rates? Coping with Ambiguity
  Using Bounded-Variation Assumptions}.
\newblock {\em The Review of Economics and Statistics\/}~{\em 100\/}(2),
  232--244.

\bibitem[\protect\citeauthoryear{Martinez-Bravo}{Martinez-Bravo}{2014}]{martinez2014role}
Martinez-Bravo, M. (2014).
\newblock The role of local officials in new democracies: Evidence from
  indonesia.
\newblock {\em American Economic Review\/}~{\em 104\/}(4), 1244--1287.

\bibitem[\protect\citeauthoryear{Masten and Poirier}{Masten and
  Poirier}{2020}]{masten2020inference}
Masten, M.~A. and A.~Poirier (2020).
\newblock Inference on breakdown frontiers.
\newblock {\em Quantitative Economics\/}~{\em 11\/}(1), 41--111.

\bibitem[\protect\citeauthoryear{Michalopoulos and Papaioannou}{Michalopoulos
  and Papaioannou}{2016}]{michalopoulos2016long}
Michalopoulos, S. and E.~Papaioannou (2016).
\newblock The long-run effects of the scramble for africa.
\newblock {\em American Economic Review\/}~{\em 106\/}(7), 1802--1848.

\bibitem[\protect\citeauthoryear{Noack and Rothe}{Noack and
  Rothe}{2024}]{noack2024bias}
Noack, C. and C.~Rothe (2024).
\newblock Bias-aware inference in fuzzy regression discontinuity designs.
\newblock {\em Econometrica\/}~{\em 92\/}(3), 687--711.

\bibitem[\protect\citeauthoryear{Rothe}{Rothe}{2017}]{rothe2017robust}
Rothe, C. (2017).
\newblock Robust confidence intervals for average treatment effects under
  limited overlap.
\newblock {\em Econometrica\/}~{\em 85\/}(2), 645--660.

\bibitem[\protect\citeauthoryear{Sanchez-Becerra}{Sanchez-Becerra}{2023}]{sanchezbecerra2023robust}
Sanchez-Becerra, A. (2023).
\newblock Robust inference for the treatment effect variance in experiments
  using machine learning.
\newblock {\em arXiv:2306.03363 [econ.EM]\/}.

\bibitem[\protect\citeauthoryear{Sasaki and Ura}{Sasaki and
  Ura}{2022}]{sasaki2022estimation}
Sasaki, Y. and T.~Ura (2022).
\newblock Estimation and inference for moments of ratios with robustness
  against large trimming bias.
\newblock {\em Econometric Theory\/}~{\em 38\/}(1), 66--112.

\bibitem[\protect\citeauthoryear{S{\l }oczy{\'n}ski}{S{\l
  }oczy{\'n}ski}{2022}]{sloczynski_interpreting_2022}
S{\l }oczy{\'n}ski, T. (2022).
\newblock Interpreting {OLS} {Estimands} {When} {Treatment} {Effects} {Are}
  {Heterogeneous}: {Smaller} {Groups} {Get} {Larger} {Weights}.
\newblock {\em The Review of Economics and Statistics\/}~{\em 104\/}(3),
  501--509.

\bibitem[\protect\citeauthoryear{Sun and Abraham}{Sun and
  Abraham}{2021}]{sun_estimating_2020}
Sun, L. and S.~Abraham (2021).
\newblock Estimating dynamic treatment effects in event studies with
  heterogeneous treatment effects.
\newblock {\em Journal of econometrics\/}~{\em 225\/}(2), 175--199.

\bibitem[\protect\citeauthoryear{Wooldridge}{Wooldridge}{2025}]{wooldridge2025twfe}
Wooldridge, J.~M. (2025).
\newblock Two-way fixed effects, the two-way mundlak regression, and
  difference-in-differences estimators.
\newblock {\em Empirical Economics\/}~{\em 69}, 2545--2587.

\bibitem[\protect\citeauthoryear{Xu}{Xu}{2018}]{xu2018costs}
Xu, G. (2018).
\newblock The costs of patronage: Evidence from the british empire.
\newblock {\em American Economic Review\/}~{\em 108\/}(11), 3170--3198.

\end{thebibliography}

\newpage
\setcounter{equation}{5}
\renewcommand{\theequation}{OA.\arabic{equation}}
\setcounter{lemma}{2}
\renewcommand{\thelemma}{OA.\arabic{lemma}}
\setcounter{claim}{0}
\renewcommand{\theclaim}{OA.\arabic{claim}}

\section*{Online Appendix A. Proofs for Section~\ref{sec:bias-variance}: Discrete-Covariate Results and No-Overlap Limit}

This section specializes to the unconfoundedness design with $W=X$ and a
discrete covariate $c(i)\in\{1,\ldots,k+1\}$, where cell 1 is the reference
cell. Let $\bar{x}_i$ be the $(k+1)\times1$ vector whose $j$th entry is
$\mathbf{1}\{c(i)=j\}$. To include an intercept without collinearity, define
$x_i=J\bar{x}_i$, where $J$ is the $(k+1)\times(k+1)$ identity matrix with its
first row replaced by a row of ones:

\[
J=\left(\begin{array}{cccc}
1 & 1 & \cdots & 1\\
0 & 1 & 0 & 0\\
\vdots & 0 & \ddots & 0\\
0 & 0 & 0 & 1
\end{array}\right),\ \bar{x}_{i}=\left(\begin{array}{c}
\mathbf{1}\{c(i) =1\}\\
\mathbf{1}\{c(i) =2\}\\
\vdots\\
\mathbf{1}\{c(i) =k+1\}
\end{array}\right).
\]

Lemma~\ref{thm:population ridge} gives a discrete-covariate representation of
the weights $a_\lambda$, and Lemma~\ref{lem:convex_weights} derives the
closed-form estimand weights in~\eqref{eq:explicit weights} for $\lambda>0$.
Lemma~\ref{lem:ridge_partial_continuity} establishes continuity of the relevant
penalized objective, and Lemma~\ref{lem:ridgeless} derives the limit of the
generalized ridge estimator as $\lambda\to0$ under lack of overlap.

\begin{lemma}\label{thm:population ridge}
For every $\lambda>0$, the generalized ridge estimand can be written as
\[
\beta_{\lambda}=\frac{\E_n\left[\left(d_{i}-p(x_i)\right)\left(1-\widetilde{x}_{i,-1}'\pi_{2, \lambda}\right)Y_{i}\right]}{\E_n\left[\left(d_{i}-p(x_i)\right)\left(1- \widetilde{x}_{i,-1}'\pi_{2, \lambda}\right)d_{i}\right]},
\]
where $p(x_j)=\E_n\left[d_i\mathbf{1}\{c(i) =c(j) \}\right] / \E_n\left[\mathbf{1}\{c(i) =c(j) \}\right]$ is the empirical propensity score, and $\pi_{2, \lambda}$
solves
\[
\min_{\pi_{2}}\ \E_n\left[\left(\left(d_{i}-p(x_i)\right)\left(1-\widetilde{x}_{i,-1}'\pi_{2}\right) \right)^{2}\right]+\lambda\E_n\left[(\widetilde{x}_{i,-1}'\pi_{2})^{2}\right].
\]
\end{lemma}
\begin{proof}[Proof for Lemma~\ref{thm:population ridge}]
By Lemma~\ref{lem:outcome ridge},
\[
\beta_{\lambda}=\frac{\E_n\left[\left(d_{i}-x_{i}'\pi_{1, \lambda}-d_{i}\widetilde{x}_{i,-1}'\pi_{2, \lambda}\right)Y_{i}\right]}{\E_n\left[\left(d_{i}-x_{i}'\pi_{1, \lambda}-d_{i}\widetilde{x}_{i,-1}'\pi_{2, \lambda}\right)d_{i}\right]},
\]
where $\pi_{1, \lambda}$ and $\pi_{2, \lambda}$ solve the penalized propensity
score regression
\[
\begin{aligned}
\min_{\pi_{1},\pi_{2}}\quad
&\E_n\left[\left(d_{i}-x_{i}'\pi_{1}-d_{i}\widetilde{x}_{i,-1}'\pi_{2}\right)^{2}\right]\\
&+\lambda\E_n\left[(\widetilde{x}_{i,-1}'\pi_{2})^{2}\right].
\end{aligned}
\]
Since $\pi_{1}$ is unpenalized and $x_i$ is based on discrete covariates,
$\pi_{1}$ can be concentrated out for any fixed $\pi_{2}$. Because
$(1-\widetilde{x}_{i,-1}'\pi_{2})$ is constant within each cell, the OLS
projection of $d_i(1-\widetilde{x}_{i,-1}'\pi_{2})$ onto $x_i$ equals
$p(x_i)(1-\widetilde{x}_{i,-1}'\pi_{2})$, which yields
\begin{align*}
&\min_{\pi_{1}}\ \E_n\left[\left(d_{i}-x_{i}'\pi_{1}-d_{i}\widetilde{x}_{i,-1}'\pi_{2}\right)^{2}\right] \\
=  &\min_{\pi_{1}}\ \E_n\left[\left(d_{i}\left(1-\widetilde{x}_{i,-1}'\pi_{2}\right) -x_{i}'\pi_{1} \right)^{2}\right] \\
= &\E_n\left[\left(\left(d_{i}-p(x_i)\right)\left(1-\widetilde{x}_{i,-1}'\pi_{2}\right) \right)^{2}\right].
\end{align*}
\end{proof}

\begin{lemma}[Closed-form weights under discrete covariates]\label{lem:convex_weights}
For every $\lambda > 0$,
the estimand in~\eqref{eq:ridge estimand}
equals~\eqref{eq:explicit weights}. In particular, the weights are
nonnegative and sum to one.
\end{lemma}

\begin{proof}
Let $\sigma^2(x_i) = p(x_i)(1-p(x_i))$ denote the conditional variance of
treatment at covariate value $x_i$.
By Lemma~\ref{thm:population ridge}, concentrating out $\pi_1$ from the
penalized propensity score objective~\eqref{eq:prop_lag} under discrete
covariates reduces it to a function of $\pi_2$ alone. The objective whose
minimizer determines the estimand weights is
\[
\E_n\bigl[\sigma^2(x_i)(1 - \widetilde{x}_{i,-1}'\pi)^2
+ \lambda(\widetilde{x}_{i,-1}'\pi)^2\bigr],
\]
using $\pi'V_x\pi = \E_n[(\widetilde{x}_{i,-1}'\pi)^2]$.
Since all observations in the same cell share the same covariate value,
$\widetilde{x}_{i,-1}'\pi$ takes a common value within each cell. In addition,
$\E_n[\widetilde{x}_{i,-1}] = 0$ implies
$\E_n[\widetilde{x}_{i,-1}'\pi] = 0$ for any $\pi$. Conversely, every
assignment of $k+1$ cell-level values satisfying this weighted-mean-zero
property is attainable. To see this, let $(r_1,\ldots,r_{k+1})$ satisfy
$\E_n[r_{c(i)}] = 0$, where $c(i) \in \{1,\ldots,k+1\}$ denotes the cell
of observation $i$. Setting $\pi_{t-1} = r_t - r_1$ for
$t = 2,\ldots,k+1$ gives $\widetilde{x}_{i,-1}'\pi = r_{c(i)}$ for all
$i$, because the $(t-1)$th component of $\widetilde{x}_{i,-1}$ equals
$\mathbf{1}\{c(i) = t\} - \E_n[\mathbf{1}\{c(i) = t\}]$ for
$t=2,\ldots,k+1$.

Since the map $\pi \mapsto (\widetilde{x}_{i,-1}'\pi)_{i=1}^n$ is
determined by the $k+1$ cell-level values and the correspondence above is a
bijection between $\pi \in \mathbb{R}^k$ and the $k$-dimensional set of
weighted-mean-zero cell-level vectors, the objective can be minimized
directly over these $k+1$ values subject to the single constraint
$\E_n[\widetilde{x}_{i,-1}'\pi] = 0$. Since each coefficient
$\sigma^2(x_i) + \lambda$ is strictly positive, the objective is strictly
convex in these cell-level values. Hence, there exists
a Lagrange multiplier $\eta_\lambda$ such that, for every observation $i$,
\[
-\sigma^2(x_i)(1 - \widetilde{x}_{i,-1}'\pi_{2,\lambda})
+ \lambda\,\widetilde{x}_{i,-1}'\pi_{2,\lambda} + \eta_\lambda = 0,
\]
which gives
\[
\widetilde{x}_{i,-1}'\pi_{2,\lambda}
= \frac{\sigma^2(x_i) - \eta_\lambda}{\sigma^2(x_i) + \lambda}.
\]
Imposing $\E_n[\widetilde{x}_{i,-1}'\pi_{2,\lambda}] = 0$ and solving
for $\eta_\lambda$ gives
\[
\eta_\lambda
= \frac{\E_n\bigl[\sigma^2(x_i)/(\sigma^2(x_i)+\lambda)\bigr]}
{\E_n\bigl[1/(\sigma^2(x_i)+\lambda)\bigr]}.
\]
This quantity is a weighted average of
$\sigma^2(x_1),\ldots,\sigma^2(x_n)$ with positive weights proportional to
$(\sigma^2(x_i)+\lambda)^{-1}$, so $\eta_\lambda \geq 0$. It follows that
\[
1 - \widetilde{x}_{i,-1}'\pi_{2,\lambda}
= \frac{\lambda + \eta_\lambda}{\sigma^2(x_i) + \lambda} > 0
\qquad (\lambda > 0),
\]
since $\lambda + \eta_\lambda > 0$ and $\sigma^2(x_i) + \lambda > 0$.
Substituting into~\eqref{eq:ridge estimand}, the weight on observation $i$'s
CATE is proportional to
$\sigma^2(x_i)(\lambda+\eta_\lambda)/(\sigma^2(x_i)+\lambda)$. The common
factor $(\lambda+\eta_\lambda)$ cancels between numerator and denominator,
yielding~\eqref{eq:explicit weights}.
\end{proof}

\begin{lemma}[Convergence as $\lambda\to0$]\label{lem:ridge_partial_continuity}
  Assume that $S(\theta)$ is quadratic and continuous in
  $\theta\in\mathbb R^{p_\theta}$, and is strongly convex with unique minimizer
  $\theta^\ast$. Assume also that $P(\theta,\psi)\ge 0$ is quadratic and continuous in
  $\theta$ and $\psi \in\mathbb R^{p_\psi}$, and $P(\theta^\ast,0)\leq M < \infty$. Define the
  penalized objective  
\[
F_\lambda(\theta,\psi)\;=\; S(\theta) \;+\; \lambda\,P(\theta,\psi),
\qquad \lambda\ge 0.
\]
Let $(\hat\theta_\lambda,\hat\psi_\lambda)$ be the minimizer of $F_\lambda$ for
$\lambda>0$. Then $\hat\theta_\lambda \to \theta^\ast$ as $\lambda\to0$.
\end{lemma}

\begin{proof}
Let $\varepsilon>0$ be arbitrary. Since $S(\theta)$ is quadratic and strongly
convex, there exists a constant $c>0$ such that
\[
S(\theta)\;\ge\; S(\theta^\ast)+c\norm{\theta-\theta^\ast}^2 .
\]
Set
\(
\xi := c\,\varepsilon^2/2.
\)
For $\lambda$ small enough so that $\lambda M\le \xi$,
\[
\inf_{(\theta,\psi)\in\mathbb R^{p_\theta+p_\psi}} F_\lambda(\theta,\psi)
\;\le\;  F_\lambda(\theta^\ast,0)
= S(\theta^\ast)+\lambda P(\theta^\ast,0)
\;\le\;
S(\theta^\ast)+\xi.
\]
Suppose $\norm{\hat\theta_\lambda-\theta^\ast}>\varepsilon$. Then
\[
F_\lambda(\hat\theta_\lambda,\hat\psi_\lambda)\;\ge\;S(\hat\theta_\lambda)
\;\ge\; S(\theta^\ast)+c\varepsilon^2
\;=\; S(\theta^\ast)+2\xi \;> \;\inf_{(\theta,\psi)\in\mathbb R^{p_\theta+p_\psi}} F_\lambda(\theta,\psi).
\]
This contradicts the definition of a minimizer. Therefore, for all sufficiently
small $\lambda$, every global minimizer
$(\hat\theta_\lambda,\hat\psi_\lambda)$ of $F_\lambda$ satisfies
$\norm{\hat\theta_\lambda-\theta^\ast}\le \varepsilon$. Since $\varepsilon>0$
was arbitrary, $\hat\theta_\lambda\to\theta^\ast$.

\end{proof}

\begin{lemma}[Behavior under no overlap]\label{lem:ridgeless}

Let $\mathcal{J}_1$ denote the set of all-treated cells
($d_i=1$ for all $i$ in the cell), and let $\mathcal{J}_0$ denote the set of
all-untreated cells ($d_i=0$ for all $i$ in the cell). Assume overlap holds in
all other cells. Then the generalized ridge estimator satisfies
\[
\lim_{\lambda \to 0} \hat{\beta}_{\lambda}
=\hat{\beta}_{\text{trim}},
\]
where $\hat{\beta}_{\text{trim}}$ is the long-regression estimator computed on
the subsample of cells with overlap, that is, the cells
$\{1,\dots,k+1\} \setminus (\mathcal{J}_0 \cup \mathcal{J}_1)$.
\end{lemma}

\Needspace{6\baselineskip}
\begin{proof}
With saturated discrete covariates, problem~\eqref{eq:gen_ridge} has the
equivalent form
\[
\min_{\bar{\gamma},\bar{\tau}} \E_n\big[(Y_i - \bar{\gamma}'\bar{x}_i - \bar{\tau}' \bar{x}_i d_i)^2 \big] + \lambda \, \E_n\big[ (\bar{\tau}' \bar{x}_i - \E_n[\bar{\tau}'\bar{x}_i])^2 \big],
\]
where $\bar{x}_i$ is the vector of cell indicators.
Under this reparameterization, the coefficient of interest is
$\hat{\beta}_{\lambda}=\E_n[\hat{\bar{\tau}}_{\lambda}'\bar{x}_i]$, where
$\hat{\bar{\tau}}_{\lambda}$ solves the ridge problem above. Let
$f_j = \E_n[\mathbf{1}\{c(i)=j\}]$ denote the cell share.

For any cell $j \in \mathcal{J}_0$ (all-untreated), $d_i = 0$ for all $i$ in the cell, so $\bar{x}_{ij} d_i = 0$.
Hence $\bar{\tau}_j$ does not enter the least-squares term.

For any cell $j \in \mathcal{J}_1$ (all-treated), $d_i = 1$ for all $i$ in the cell, so $\bar{x}_{ij} d_i = \bar{x}_{ij}$.
Writing $\tilde{\bar{\gamma}}_j = \bar{\gamma}_j + \bar{\tau}_j$, $\bar{\tau}_j$ again does not enter the squared-error term separately.

Let $\mathcal{N} = \mathcal{J}_0 \cup \mathcal{J}_1$ be the set of cells without overlap.
By Lemma~\ref{lem:ridge_partial_continuity}, as $\lambda \to 0$, the
coefficients $\hat{\bar{\tau}}_{k,\lambda}$ for $k \notin \mathcal{N}$ converge
to the OLS coefficients that minimize
\begin{equation}
\min_{\bar{\gamma}, \bar{\tau}_{k \notin \mathcal{N}}} \E_n \left[ \left( Y_i - \bar{\gamma}' \bar{x}_i - \sum_{k \notin \mathcal{N}}\bar{\tau}_k \bar{x}_{ik} d_i \right)^2 \right].\label{eq:ridge limit}
\end{equation}
The OLS problem~\eqref{eq:ridge limit} gives
\begin{align*}
    \hat{\bar{\tau}}_k
    &= \frac{\E_n[Y_i \bar{x}_{ik} d_i]}{\E_n[\bar{x}_{ik} d_i]}
    -\frac{\E_n[Y_i \bar{x}_{ik} (1-d_i)]}{\E_n[\bar{x}_{ik} (1-d_i)]},
    && k \notin \mathcal{N},  \\
    \hat{\bar{\gamma}}_k
    &= \frac{\E_n[Y_i \bar{x}_{ik} (1-d_i)]}{\E_n[\bar{x}_{ik} (1-d_i)]},
    && k \notin \mathcal{N},  \\
    \hat{\bar{\gamma}}_j
    &= \frac{\E_n[Y_i \bar{x}_{ij}]}{\E_n[\bar{x}_{ij}]},
    && j \in \mathcal{N}.
\end{align*}

For any $j \in \mathcal{N}$, the coefficient $\bar{\tau}_j$ does not enter the squared-error term,
so the ridge objective depends on $\bar{\tau}_j$ only through the penalty
$\E_n[(\bar{\tau}'\bar{x}_i - \E_n[\bar{\tau}'\bar{x}_i])^2]$.
Define $\sigma_{jk} := \E_n[\bar{x}_{ij}\bar{x}_{ik}]
- \E_n[\bar{x}_{ij}]\E_n[\bar{x}_{ik}]$, the sample covariance
between cell indicators $j$ and $k$. The penalty can then be expanded as
$\sum_{j,k} \bar{\tau}_j \bar{\tau}_k \sigma_{jk}$.
Taking the derivative with respect to $\bar{\tau}_j$ for $j \in \mathcal{N}$ gives the linear system:
\[
\bar{\tau}_j\,\sigma_{jj}
+ \sum_{k \in \mathcal{N},\, k\neq j}
\bar{\tau}_k\,\sigma_{jk}
= -\sum_{k \notin \mathcal{N}} \bar{\tau}_k\,\sigma_{jk},
\qquad j \in \mathcal{N}.
\]
For categorical indicators, $\sigma_{jj} = f_j(1-f_j)$ and
$\sigma_{jk} = -f_j f_k$ for $j \neq k$.
Substituting
$\lim_{\lambda \to 0} \widehat{\bar{\tau}}_{k,\lambda}
= \hat{\bar{\tau}}_k$ for $k \notin \mathcal{N}$ into this system and solving
the first-order conditions yields
\[
\lim_{\lambda \to 0} \hat{\bar{\tau}}_{j,\lambda} = \frac{\sum_{k \notin \mathcal{N}} f_k \hat{\bar{\tau}}_k}{1 - \sum_{k \in \mathcal{N}} f_k},
\quad \text{for all }j \in \mathcal{N}.
\]

Finally, by linearity of expectation,
\begin{align*}
\lim_{\lambda \to 0} \hat{\beta}_\lambda
&= \lim_{\lambda \to 0} \E_n[\hat{\bar{\tau}}_\lambda' \bar{x}_i]
= \E_n[(\lim_{\lambda \to 0} \hat{\bar{\tau}}_\lambda)' \bar{x}_i] \\
&= \sum_{k \notin \mathcal{N}} f_k \hat{\bar{\tau}}_k
+ \sum_{j \in \mathcal{N}} f_j \cdot \lim_{\lambda \to 0}\hat{\bar{\tau}}_{j,\lambda} = \frac{\sum_{k \notin \mathcal{N}} f_k \hat{\bar{\tau}}_k}{\sum_{k \notin \mathcal{N}} f_k} = \hat{\beta}_{\mathrm{trim}},
\end{align*}
which is the long regression coefficient in the sample restricted to cells with overlap.
\end{proof}
\end{document}